\documentclass[traditabstract]{aa} 
\usepackage{longtable}

\usepackage{graphicx}
\usepackage{supertabular}

\begin{document}
   \title{Studying the kinematics of the giant star-forming region \\30 Doradus\thanks{Based on observations made with ESO Telescopes at the La Silla Paranal Observatory under programmes 072.C-0348 and 182.D-0222.} \subtitle{I. The data}}

   \author{S. Torres-Flores\inst{1}, 
                R. Barb\'a\inst{1,2}, 
                J. Ma\'iz Apell\'aniz\inst{3}, 
                M. Rubio\inst{4}, 
                G. Bosch\inst{5},
                V. H\'enault-Brunet\inst{6},
                C. J. Evans\inst{7}
                }

   \institute{Departamento de F\'isica, Universidad de La Serena, Av. Cisternas 1200 Norte, La Serena, Chile
         \and
             Instituto de Ciencias Astron\'omicas, de la Tierra y del Espacio, Casilla 467, 5400 San Juan, Argentina
          \and   
             Instituto de Astrof\'isica de Andaluc\'ia-CSIC, Glorieta de la Astronom\'ia s/n, 18008 Granada, Spain
          \and
             Departamento de Astronom\'ia, Universidad de Chile, Casilla 36-D, Santiago, Chile
          \and   
             Facultad de Ciencias Astron\'omicas y Geof\'isicas, Universidad Nacional de la La Plata, Paseo del Bosque s/n, 1900 La Plata, Argentina 
           \and
           Institute for Astronomy, Royal Observatory Edinburgh, Blackford Hill, Edinburgh, EH9 3HJ, UK
           \and
           UK Astronomy Technology Centre, Royal Observatory Edinburgh, Blackford Hill, Edinburgh, EH9 3HJ, UK    
             }

   \date{Received ; accepted }

 
  \abstract
  {We present high-quality VLT-FLAMES optical spectroscopy of the
    nebular gas in the giant star-forming region 30~Doradus.  In this
    paper, the first of a series, we introduce our observations and
    discuss the main kinematic features of 30~Dor, as revealed by the
    spectroscopy of the ionized gas in the region. The primary data
    set consists of regular grid of nebular observations, which we
    used to produce a spectroscopic datacube of 30~Dor, centered on the
    massive star cluster R136 and covering a field-of-view of
    10$'$\,$\times$\,10$'$.  The main emission lines present in the
    datacube are from H$\alpha$ and [N{\sc ii}]
    $\lambda\lambda$6548,\,6584. The H$\alpha$ emission-line profile
    varies across the region from simple single-peaked emission to
    complex, multiple-component profiles, suggesting that
    different physical mechanisms are acting on the excited gas.  To
    analyse the gas kinematics we fit Gaussian profiles to the
    observed H$\alpha$ features.  Unexpectedly, the narrowest
    H$\alpha$ profile in our sample lies close to the supernova
    remnant 30~Dor~B. We present maps of the velocity field and
    velocity dispersion across 30~Dor, finding five previously
    unclassified expanding structures. These maps highlight the
    kinematic richness of 30~Dor (e.g. supersonic motions), which will
    be analysed in future papers.





}
{}{}{}{}
\authorrunning{S. Torres-Flores et al.}

   \maketitle
%

\section{Introduction}
\label{introduction}

Giant H\,{\sc ii} regions (GHRs) are known as the site of massive
young clusters which are rich in massive stars. The strong stellar
winds and the evolution of these massive stars can disrupt the
interstellar medium of these GHRs, resulting in the formation of
expanding superbubbles. In the simplest scenario, the creation of
superbubbles can be done by scaling the kinetic energy luminosity from
the single-star models of Weaver et al. (1977). These models assume
that shocked stellar winds dominate the dynamics of these bubbles.
However, analysis of several superbubbles created by OB associations
in the Large Magellanic Cloud (LMC) (Oey 1996) and some extragalactic
GHRs (Ma\'iz-Apell\'aniz \& Walborn 2001, MacKenty et al. 2000)
reveals that such a model can not explain the observations.  Recently,
Lopez et al. (2011) studied the different feedback processes that are
taking place in the 30~Doradus region in the LMC. They suggested that
stellar feedback is the main process responsible in shaping the
expansion of this GHR. In a different approach, Pellegrini et al.
(2011) found that the large-scale structure of 30~Dor is dominated by
a system of X-ray bubbles, which are in equilibrium between them. In
spite of several studies to investigate the origin of bubbles in GHRs,
there is no consensus as to the mechanisms responsible, even in the
well-studied case of 30~Dor.

A related, unsolved problem regarding GHRs is the origin of supersonic
velocities seen in spatially-integrated nebular profiles. Given that
supersonic motions should rapidly be dissipated, an energy source is
needed to maintain them. Three possible sources exist: kinetic energy
input from stellar winds and supernova explosions (SN), gravity, and
photoionization. All of them can supply enough power to maintain the
supersonic velocity dispersion but it is not clear which one(s) is/are
responsible (Chu \& Kennicutt 1994, Tenorio-Tagle et al. 1993, 1996,
Melnick et al. 1999). High-resolution spectroscopic data of nearby
GHRs is necessary to disentangle these phenomena through the study of
the kinematics of the warm gas. In this context, 30~Dor is one of the
closest and most appropriate targets for undertaking a more detailed study.

The 30~Dor nebula is the largest H\,{\sc ii} region in the Local Group
and the most powerful source of H$\alpha$ emission in the LMC.  At its
core is the young massive cluster R136, host to a very large
concentration of massive hot and luminous stars (see Crowther et al.
2010). Given all the information that can be extracted from 30~Dor,
this object has been called the starburst ``Rosetta Stone'' (Walborn
1991).  Due to its nature, 30~Dor has been the target of a range of
multiwavelength studies (infrared, optical, ultraviolet and X-ray
data) to try to disentangle its structure -- both in
terms of its stellar populations and the distribution of gas and dust.
For instance, Bosch et al. (2009) searched for massive binary stars in
the ionizing cluster of 30~Dor, finding a binary candidate rate of
50$\%$. These authors also derived the binary-corrected, virial mass
of the cluster, which corresponds to 4.5$\times$10$^{5}M_{\odot}$,
suggesting that it could be a candidate for a future globular cluster.
More recently, the VLT-FLAMES Tarantula Survey (VTFS, Evans et al.
2011) has opened a new window in the study of massive binary systems
in 30~Dor (Sana et al. 2012), via the large number (over 800) of stars
for which there are now high-quality spectroscopic data. 

From imaging with the \textit{Hubble Space Telescope}, Walborn,
Ma\'{\i}z Apell\'aniz \& Barb\'a (2002) found some wind-blown cavities
and several filamentary structures.  While at X-ray wavelengths, Wang
(1999) and Townsley et al. (2006a, 2006b) found several X-ray bubbles
and point sources in 30~Dor. In the case of the diffuse X-ray
structures, which are spatially associated with high-velocity, optical
emission-line clouds (Chu \& Kennicutt 1994), Townsley et al.  (2006a)
found that not every high-velocity feature displays bright X-ray
emission. In this sense, optical kinematical analysis of the bright
X-ray regions is needed in order to understand this phenomenon.

The kinematics of the warm ionized gas in 30~Dor has also been well
studied. For example, Smith \& Weedman (1972) used a Fabry-Perot
instrument to map the [O{\sc iii}] $\lambda$5007 emission line, and
suggested that the fast internal motions in the region are produced
by the winds of WR stars. Chu \& Kennicutt (1994) and Melnick et al.
(1999) both studied the kinematics of the ionized gas in 30~Dor, using
echelle and long-slit spectroscopy, respectively.  Chu \& Kennicutt
(1994) found that 30~Dor displays very complex kinematics, with
several fast expanding shells (that are coincident with extended X-ray
sources) that can not be explained by stellar-wind models, and they suggested
that SN remnants can solve that problem. Melnick et al. (1999) found
complex H$\alpha$ profiles in several regions of 30~Dor. They used
their observations to study the mechanisms responsible for the
line-broadening observed in the emission lines of H\,{\sc ii} regions
and found a low-intensity broad component that explains the wings in
the integrated profile of 30~Dor. Lastly, Redman et al. (2003) used
echelle spectroscopy to study the gas kinematics in the outskirts of
30~Dor. These authors found some high-speed features, that they
interpreted as shells formed by stellar winds and SNe.

\begin{figure}[h!]
\centering
\includegraphics[width=0.4\textwidth]{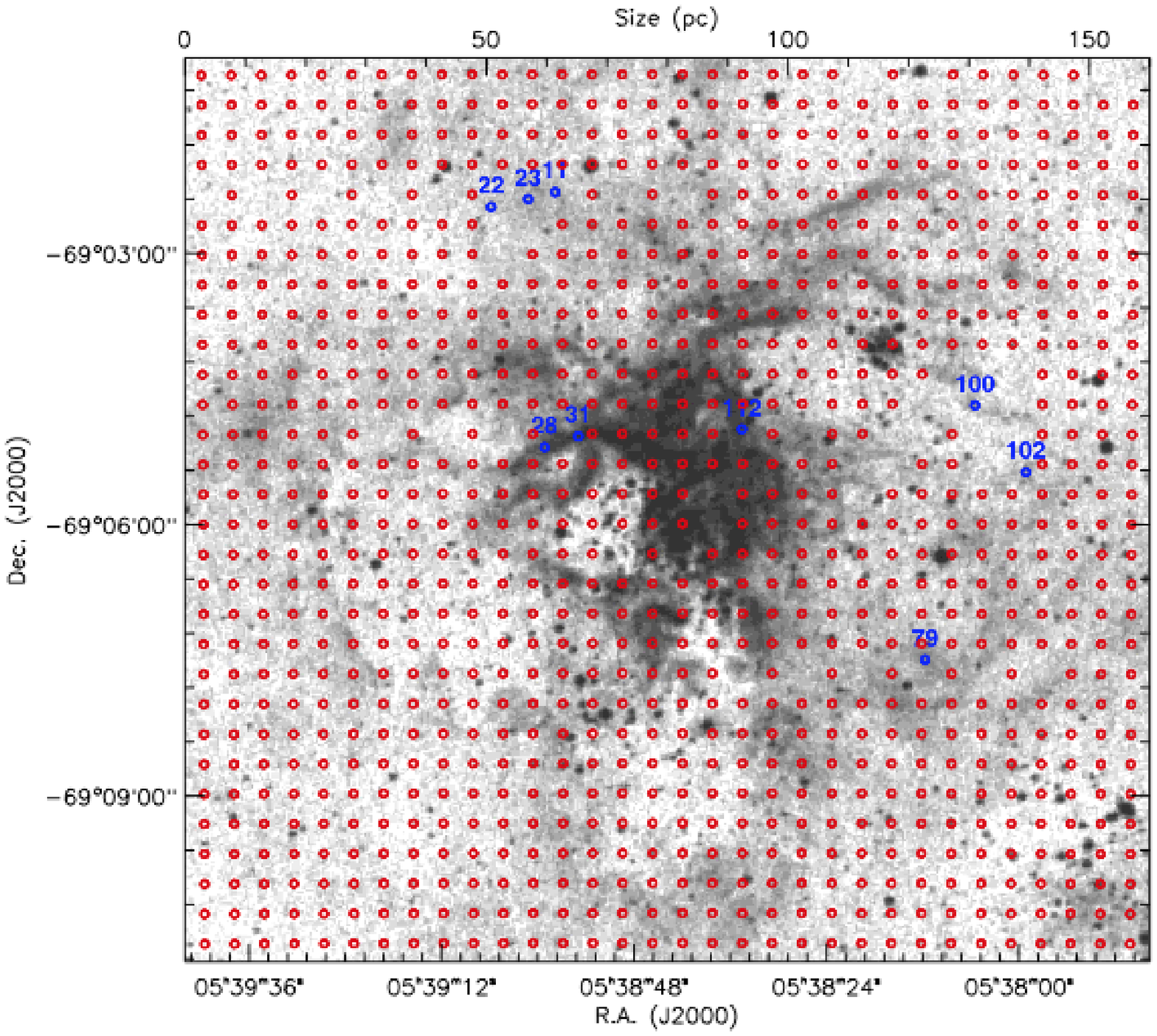}
\includegraphics[width=0.4\textwidth]{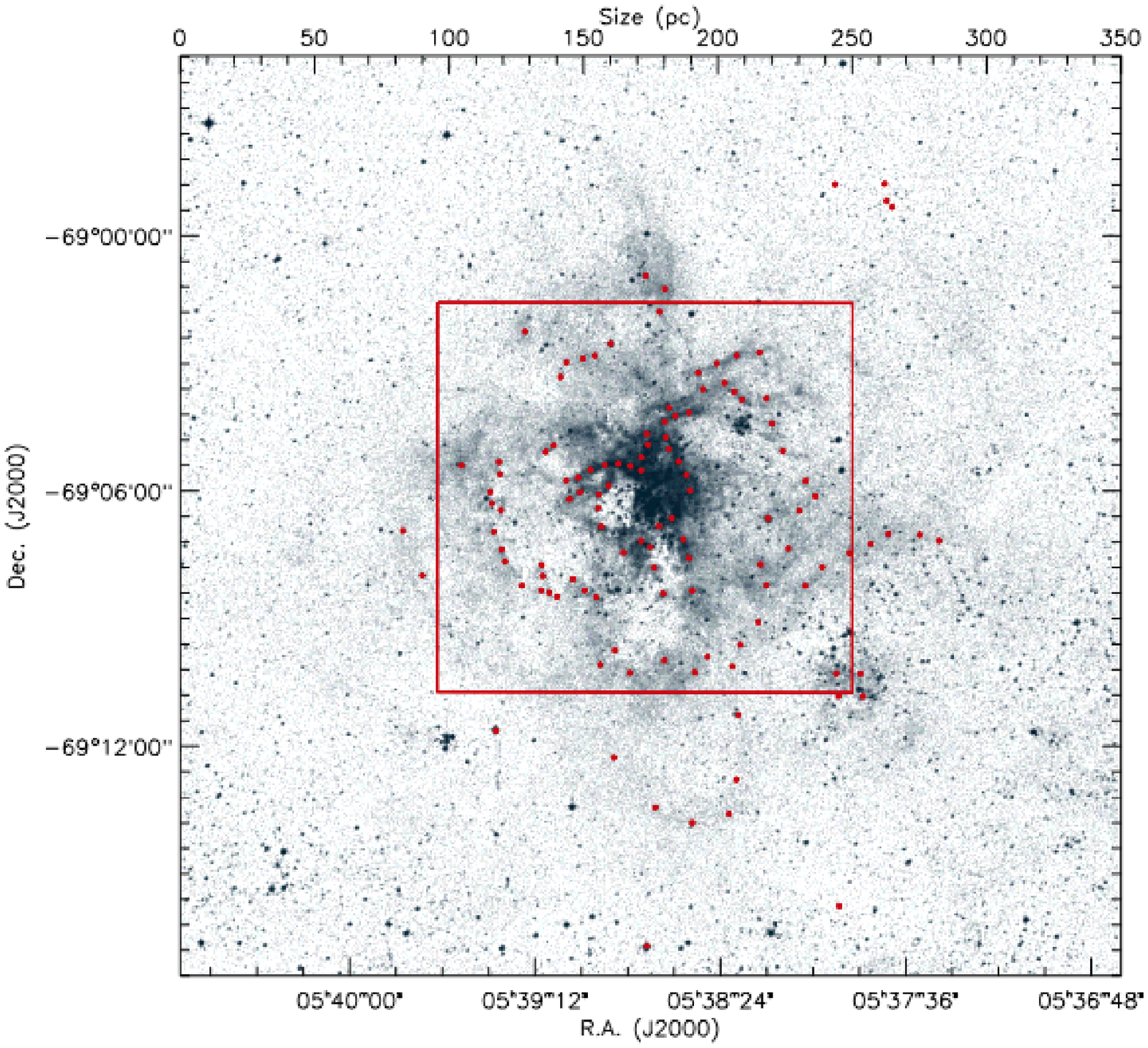}
\includegraphics[width=0.4\textwidth]{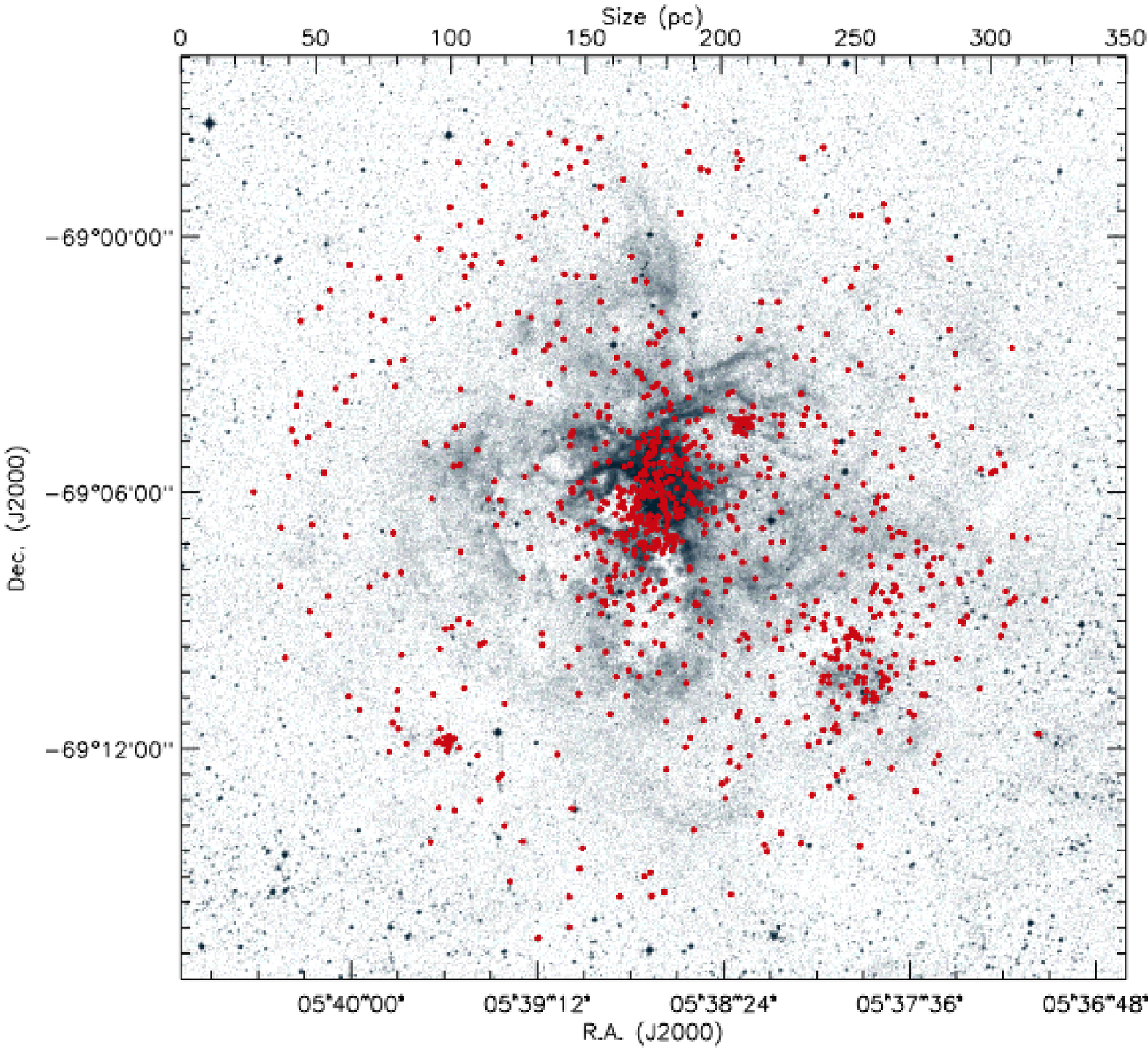}
\caption{Top panel: GIRAFFE/MEDUSA regular grid overlaid on an optical Digital Sky Survey
  image of 30~Dor. The spatial separation between each fiber corresponds to 20\arcsec ($\sim$5 pc). Middle panel: MEDUSA irregular grid. Bottom panel:
  MEDUSA stellar grid (VFTS survey). The field-of-view on the sky
  shown in the lower two panels is larger than in the top panel to
  show all the fibers available. The red rectangle in the middle panel
  indicates the region covered by the regular grid.}
\label{pointing}
\end{figure}

These studies have each provided useful insights about the kinematics
of 30~Dor, but the origin of superbubbles and the supersonic velocity
dispersions in GHRs are still open questions. In fact, Melnick et al.
(1999) suggested that it would be crucial to study the complete 30~Dor
nebula at high spatial resolution. Thus, to address these problems, we
have obtained high-resolution spectroscopy of 30~Dor, using the
FLAMES--Giraffe spectrograph on the Very Large Telescope (VLT) at the
Paranal Observatory.  Preliminary results from this programme have
been published by Torres-Flores et al. (2011). 

In this paper we present the data and discuss the general kinematics
of 30~Dor, based on analysis of its nebular emission-lines. In future
papers we will present analysis of: i) the width of the different
emission lines and their contribution to the integrated emission-line
profile of 30~Dor; ii) the kinematics of specific regions; iii) the
wings detected in the integrated H$\alpha$ profile of 30~Dor; iv) the
connection between the kinematics of the ionized warm gas and the
X-ray emission in this GHR. This paper is organized as follows. In \S
\ref{observations} we present the data and the data-reduction process.
In \S \ref{analysis} we present the data analysis. In \S \ref{results}
we show the general results obtained with the current data set.
Finally, in \S \ref{summary} we summarize our findings.

\section{Observations}
\label{observations}

\subsection{Data sets and fiber positioning}
\label{datasets}

To study the detailed kinematics of the ionized gas in 30~Dor we used
the MEDUSA fiber mode of the VLT FLAMES--Giraffe spectrograph
(Pasquini et al. 2002).  Given the proximity of 30~Dor, FLAMES allows
us to map, at high spectral and spatial resolution and high
signal-to-noise, the main nebular emission lines from the gas, i.e.,
H$\alpha$ and [N{\sc ii}] at 6548 and 6584\,{\AA}. In this regard, the
multiplex of the MEDUSA observing mode enables us to sample a large
number of points simultaneously across a relatively wide
field-of-view; crucial when studying an extended region like 30~Dor.
Given the complex morphology of 30~Dor, three different data sets are
used to analyse the its kinematic features: a regular nebular 
grid, an irregular nebular grid (where the fibers were located in the
brightest regions of 30~Dor), and finally a stellar grid. In the
following we describe each data set in detail.

\begin{table*}[t!]
\begin{center}
\begin{minipage}[t]{\columnwidth}
\caption{VLT-FLAMES/MEDUSA observing log for the regular nebular grid.}
\renewcommand{\footnoterule}{}  
\begin{tabular}{ccccc}
\hline \hline
Field & Coordinates& Setup &  N. Fibers &Exp. [s]\\
\hline
1  &  05:38:42.0, -69:05:40.0 & HR14, Medusa1  & 101 & 2$\times$600 \\ 
2  &  05:38:45.7, -69:05:40.0 & HR14, Medusa1  & 101 & 2$\times$600 \\
3  &  05:38:45.7, -69:05:20.0 & HR14, Medusa1  & 76 & 2$\times$600 \\
4  &  05:38:42.0, -69:05:20.0 & HR14, Medusa1  & 101 & 2$\times$600 \\
5  &  05:38:42.0, -69:05:40.0 & HR14, Medusa2  & 98 & 2$\times$600 \\ 
6  &  05:38:45.7, -69:05:40.0 & HR14, Medusa2  & 98 & 2$\times$600 \\
7  &  05:38:45.7, -69:05:20.0 & HR14, Medusa2  & 98 & 2$\times$600 \\
8  &  05:38:42.0, -69:05:20.0 & HR14, Medusa2  & 98 & 2$\times$600 \\
9  &  05:38:42.0, -69:09:00.0 &  HR14, Medusa1  & 71 & 2$\times$600 \\
10  &  05:38:42.0, -69:08:40.0 &  HR14, Medusa1 & 69 & 2$\times$600 \\
\hline
\label{obslogregular}
\end{tabular}
\end{minipage}
\end{center}
\end{table*}

\subsubsection{The regular grid}
\label{regulargrid}

The observations of the regular nebular grid were carried out on
December 3, 2003, using the HR14A Giraffe setting. This provided
spectral coverage ranging from 6300 to 6691\,{\AA}, with a notional spectral
resolving power of $R$\,=\,17\,740 at the central wavelength. We covered a field-of-view of
10$'$\,$\times$\,10$'$ centered on R136. Given that we were limited by
the MEDUSA fiber separations, we used three different fiber
configurations in offset positions to adequately sample our
field-of-view; a total of 10 fields were observed, as summarised in
Table~\ref{obslogregular}.  These fields give a regular grid of
32\,$\times$\,30 positions with a spatial sampling of 20$''$, as shown
in the top panel of Fig.~\ref{pointing}, where the position of each
fiber is indicated by a red circle. The aperture of each MEDUSA fiber 
corresponds to 1.2$''$ on the sky ($\sim$0.3 pc).

Our configurations did not cover some positions given that two MEDUSA
fibers were broken, producing some voids in the spatial sampling (see
Fig.~\ref{pointing}). Out of 960 fiber positions, 49 positions in the
grid were not observed. We used nine fibers of the irregular grid to
cover some of these missing positions (fibers nos. 11, 22, 23, 28, 31,
79, 100, 102 and 112). The positions of these fibers are indicated by
blue numbered circles in the top panel of Fig.~\ref{pointing}. In the
case of fibers 28 and 31, we calculated an average of the two
spectra and replaced this value at the corresponding missing position.
The remaining 41 missing positions were filled with the average of the
eight closest (and spatially linked) spectra. In this sense, we
caution the reader that the spectra visible in the positions of the
missing fibers do not represent the real emission at these locations
(see Fig.~\ref{pointing}).

The centers of the different configurations, setups, and exposure
times used to observe 30~Dor are listed in Table~\ref{obslogregular}.
Given the high surface-brightness displayed by 30~Dor, two sets of
observations were taken to avoid saturation effects. In the first
instance we took three exposures of 60\,s and then we took two
exposures of 600\,s.  At the end, we co-added the two observations of
600\,s. The fiber positions for the regular grid are listed in
Table~\ref{table_regular_all} of the Appendix; when a fiber was taken
from the irregular grid (as described above), it is labelled as `IG'.

\begin{figure*}[t!]
\centering
\includegraphics[width=0.8\textwidth]{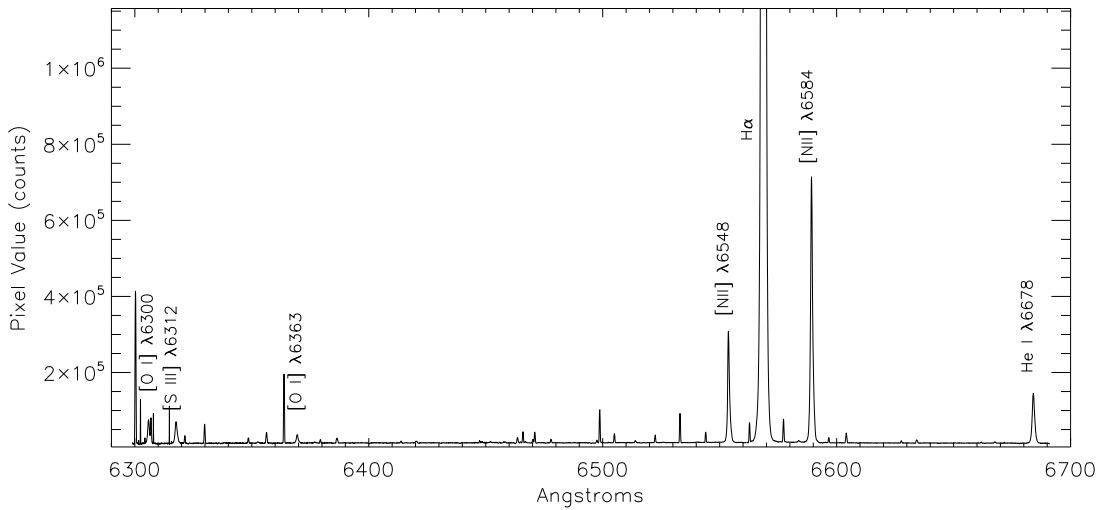}
\caption{Integrated nebular spectrum of 30~Dor obtained from summing
  all of the spectra from the regular grid. The most intense emission
  line is H$\alpha$, but the intensity axis in the plot is truncated
  to show the low-intensity emission lines.}
\label{espectro_total_integrado}
\end{figure*}

Thus, from the regular grid we have obtained 911 equally-spaced
spectra across an area of 10'\,$\times$\,10' centered on R136. With that data in-hand, we sorted all the spectra in right ascension and declination. Given that each fiber gave us spectral information for each equally-spaced point, the regular grid enabled us to produce a high-resolution spectroscopic datacube of 30~Dor, spanning 6300 to 6691\,\AA. This datacube can provide us spatial and spectral information simultaneously, where the spatial information depends on the fiber positions and the spectral information depends on the wavelengths covered by the spectra. For instance, a slice of this datacube centered at H$\alpha$ can give us a monochromatic image of 30~Dor, albeit limited in spatial resolution to 20$''$ by our fiber separation. Despite this fact, the datacube give us an excellent means by which to investigate the kinematics of the large expanding structures in
30~Dor. To illustrate the spectral coverage and primary nebular
emission features in the spectra, the integrated spectrum (summed from
all spectra in the datacube) is shown in
Fig.~\ref{espectro_total_integrado}. Given that the most intense emission lines in Fig.~\ref{espectro_total_integrado} correspond to H$\alpha$ and [N{\sc ii}] $\lambda$6584, we have derived two sub-datacubes centered in these lines, where these sub-cubes cover a range of 500 km s$^{-1}$ in velocity.

In this paper we focus on analysis of the regular grid, but we now
briefly introduce the other two data sets as they will be employed in the
future studies.

\subsubsection{The irregular grid}
\label{irregulargrid}

Additional observations were obtained in the same observing run as the
regular grid. This `irregular grid' of MEDUSA positions was selected
to observe the brightest nebular regions of 30~Dor, which can be
associated with photodissociation regions which lie between the
bright nebular and molecular phases. The same high-resolution setup
(i.e. HR14A) was used as the regular grid.  To reach a high
signal-to-noise ratio, nine exposures (each of 600\,s) were obtained
with this configuration; the positions are shown in the middle panel
of Fig.~\ref{pointing} (red circles). The red rectangle indicates the
field-of-view covered by the regular nebular grid; as can be seen most
of the fibers were located across the filamentary structure of 30~Dor.
Each fibre position is listed in Table~\ref{table_irregular_all} of
the Appendix.

\subsubsection{The stellar grid}
\label{stellargrid}

The observations of the stellar grid were obtained as part of the VFTS
(Evans et al. 2011). This multi-epoch spectral survey aims to detect
massive binary systems and determine their nature and evolution. The
VFTS also aims to study the properties of stellar winds and rotational
mixing in O-type stars. To carry out these studies, the survey has
observed 800 massive stars, as shown in the bottom panel of
Fig.~\ref{pointing}. Of relevance here, the red-optical VFTS data were
observed with the HR15N Giraffe setup, which has a spectral coverage
ranging from 6442 to 6817\,{\AA}, at a spectral resolving power of
$R$\,$=$\,16\,000.  The HR15N setup provides the additional nebular
diagnostic of the [S{\sc ii}] lines at $\lambda\lambda$6716,
6731\,\AA, which are not covered by the HR14A observations.  Given the
large spatial coverage of the VFTS survey, these data can provide
additional insights in the study of the kinematics of 30~Dor.

\subsection{Data reduction}
\label{datareduction}

Reduction of the 2003 data was performed with the ESO pipeline GASGANO and
EsoRex software. We observed three bias and flats, which were combined
to correct our observations. The wavelength calibration was performed
by using the ThAr calibration lamp, from which the instrumental
resolution was measured to have a FWHM of 0.4\,\AA, which corresponds
to $R$\,$=$\,16\,400 at H$\alpha$. All the spectra were corrected to the heliocentric rest frame.

We have compared our processed data with the data available in the
Giraffe archive\footnote{http://giraffe-archive.obspm.fr/}. By visual
inspection, the two reductions are in good agreement.  Cosmic rays
were removed using the task {\sc crreject} in {\sc iraf}. Once the
data were bias/flat-field corrected and wavelength calibrated, we
removed the continuum emission present in the spectra by fitting a
polynomial (using the {\sc iraf} {\sc continuum} task). In a few cases
in the regular grid, the fibers lie close to some stars; for these we
use a high-order polynomial to remove the continuum emission.  To
remove the sky emission lines, we fit Gaussians to one of the lowest
intensity spectra. This allowed us to create a template of sky lines,
which were removed from all the spectra.

\begin{figure*}
\centering
\includegraphics[width=\textwidth]{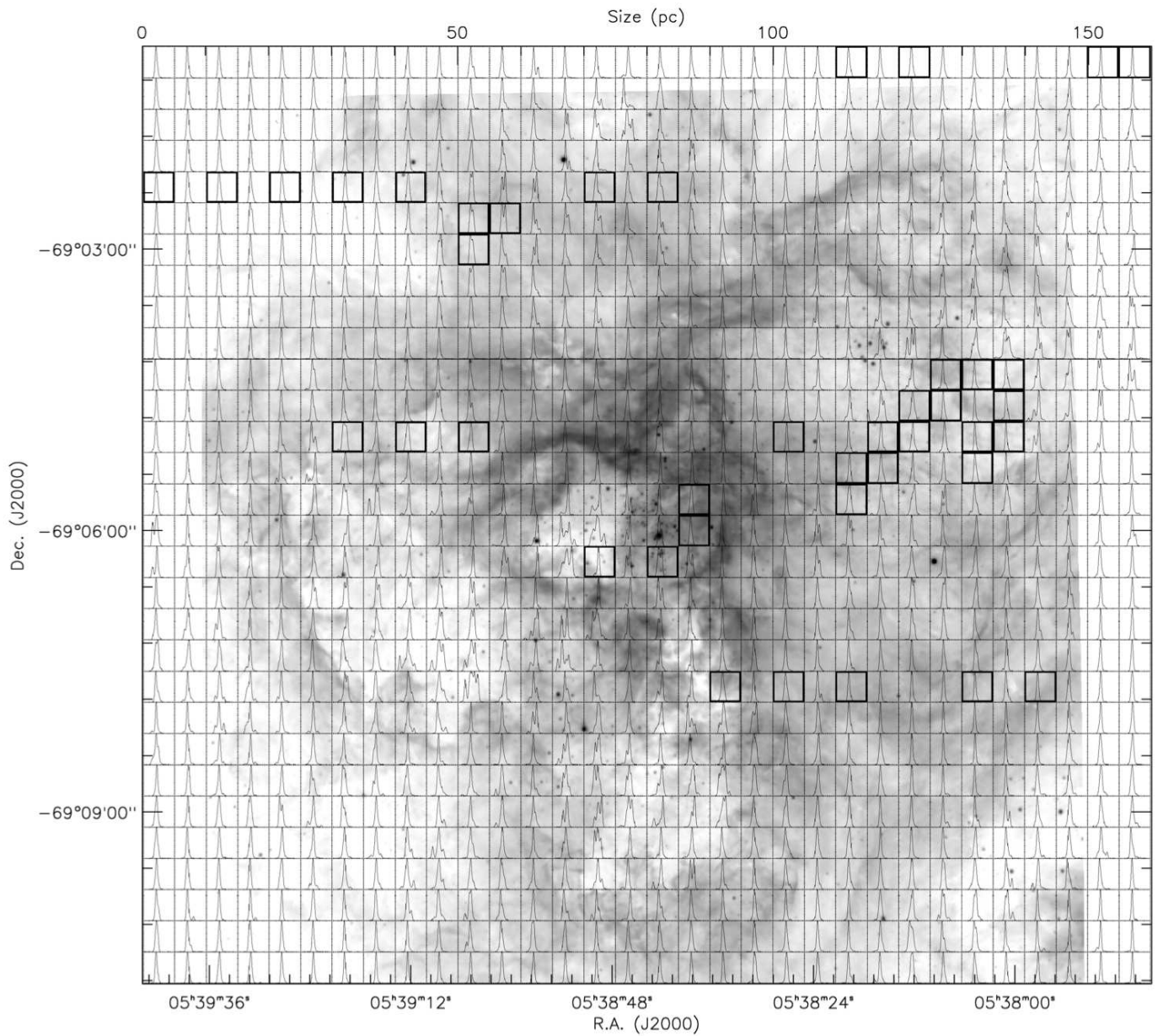}
\caption{ESO/La Silla H$\alpha$ image of 30~Dor (taken with the EMMI instrument), on which we have
  over-plotted the H$\alpha$ profile from each fiber of the regular
  nebular grid. Each small box is 20$''$\,$\times$\,20$''$ (5 pc $\times$ 5 pc), with
  the fibers located at the center of each box. Thick boxes indicate the positions for which we have interpolated the spectra (see \S 2.1.1).}
\label{plot_adhoc_cut_Halpha_image}
\end{figure*}

\section{Analysis}
\label{analysis}

Past efforts to understand the internal kinematics of H\,{\sc ii}
regions have employed profile widths of their emission lines as a
diagnostic (e.g., Chu \& Kennicutt 1994). Following a similar
approach, we use the H$\alpha$ line from our observations to
characterize the kinematics of the gas, as it presents the highest
signal-to-noise ratio and can be easily resolved into different
components in our data. To obtain the line widths we fitted Gaussians
automatically to each observed H$\alpha$ profile. In several cases a
single Gaussian does not represent the real shape of the multiple
profiles that can be seen in this star-forming region (e.g., Melnick
et al. 1999). However, this exercise provides some insight about the
general behaviour of the H$\alpha$ emission.

We used the {\sc mpfit} package in {\sc idl} (Markwardt 2009) to obtain the center, peak intensity and the observed width
($\sigma_{obs}$, uncorrected by instrumental and thermal broadening)
of each emission line. Thereafter, where relevant, we fitted multiple
Gaussians to the observed profiles using the {\sc pan}
package\footnote{http://ifs.wikidot.com/pan}, which enabled us to
select the number of components for each observed profile\footnote{For
  consistency, we have checked the results derived from {\sc mpfit}
  and {\sc pan} in the case when just one Gaussian is fitted, finding
  the same results.}. This kind of analysis is necessary given the
complexity of the emission lines in 30~Dor. For example, Melnick et
al. (1999) showed that more than three Gaussian components are needed
to reproduce some of the observed profiles. We note that this kind of
exercise is (to some degree) arbitrary, given that in several cases we
can not resolve two superposed H$\alpha$ profiles (which lie at the
same radial velocity, for example), which we would fit with just one
Gaussian.

The width of the observed H$\alpha$ profiles take into account the
instrumental ($\sigma_{in}$) and thermal ($\sigma_{th}$) widths. The
former value depends on the resolution of the instrument, while the
latter depends on the thermal motions of the Hydrogen. To obtain a corrected value for the width of the H$\alpha$ emission line, we have
subtracted $\sigma_{in}$ and $\sigma_{th}$ from $\sigma_{obs}$ as
follows:
$\sigma^{2}=\sigma_{obs}^{2}-\sigma_{in}^{2}-\sigma_{th}^{2}$, where
$\sigma$ represents the true width of the H$\alpha$ line. As discussed
in \S\ref{datareduction}, $\sigma_{in}$ was derived from the
calibration lamp exposures and has a value of
$\sigma_{in}$\,$=$\,7.8\,km\,s$^{-1}$. For $\sigma_{th}$ we assume
hydrogen gas at an electronic temperature of 10$^{4}$\,K, from which
$\sigma_{th}$\,$\sim$\,9.1\,km\,s$^{-1}$.

\begin{figure*}[t!]
\centering
\includegraphics[width=0.85\textwidth]{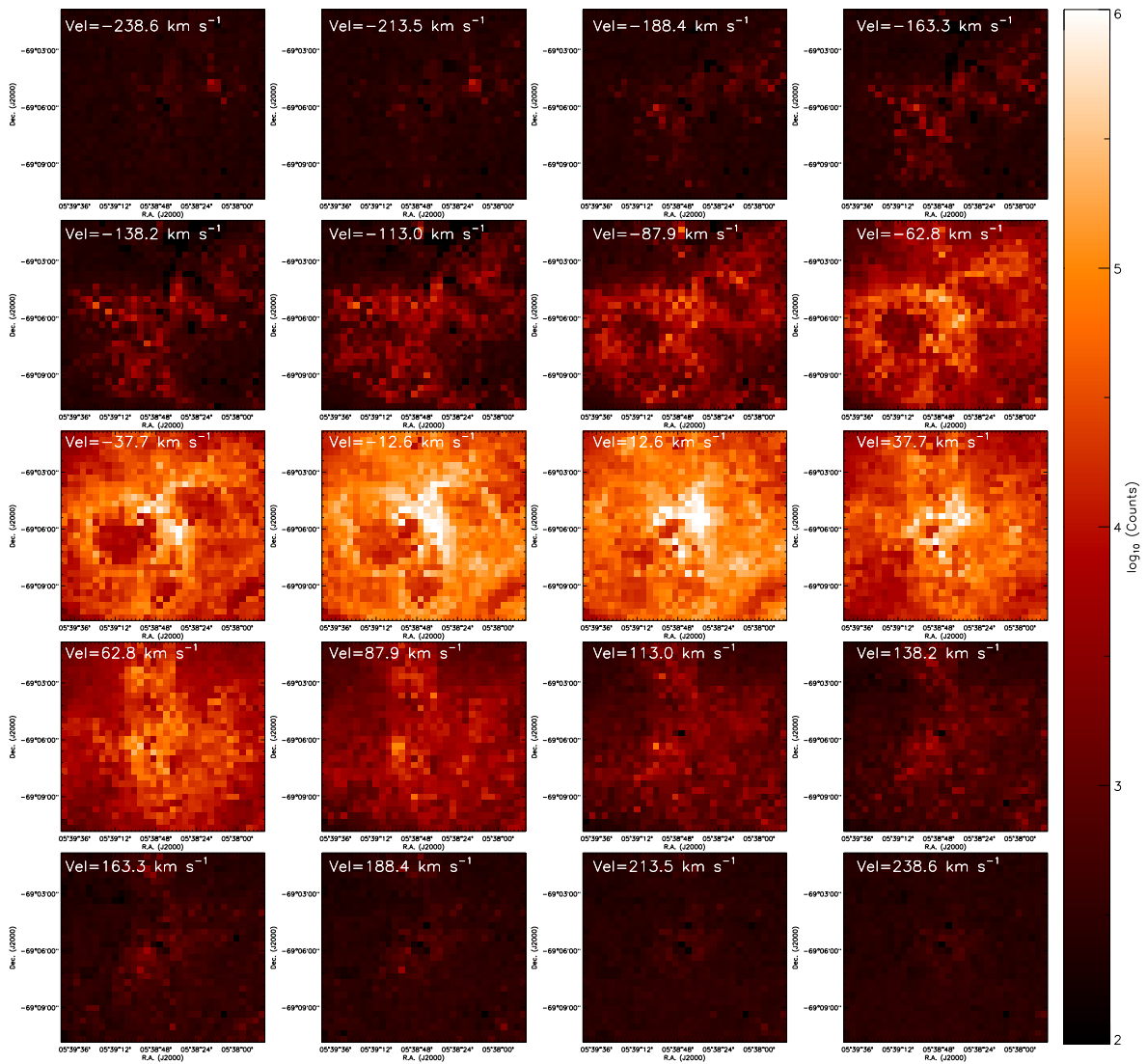}
\caption{Velocity-sliced view of the 30~Dor datacube
  centered on the H$\alpha$ line. The velocity with
  respect to the center of the H$\alpha$ line is shown in the upper left-hand corner of each panel (negative and positive velocities imply radially approaching and receding velocities, respectively).
  In this figure, a logarithmic scale is used for the intensity of
  each map.}
\label{HI_frames_Halpha}
\end{figure*}

\begin{figure*}[t!]
\centering
\includegraphics[width=0.85\textwidth]{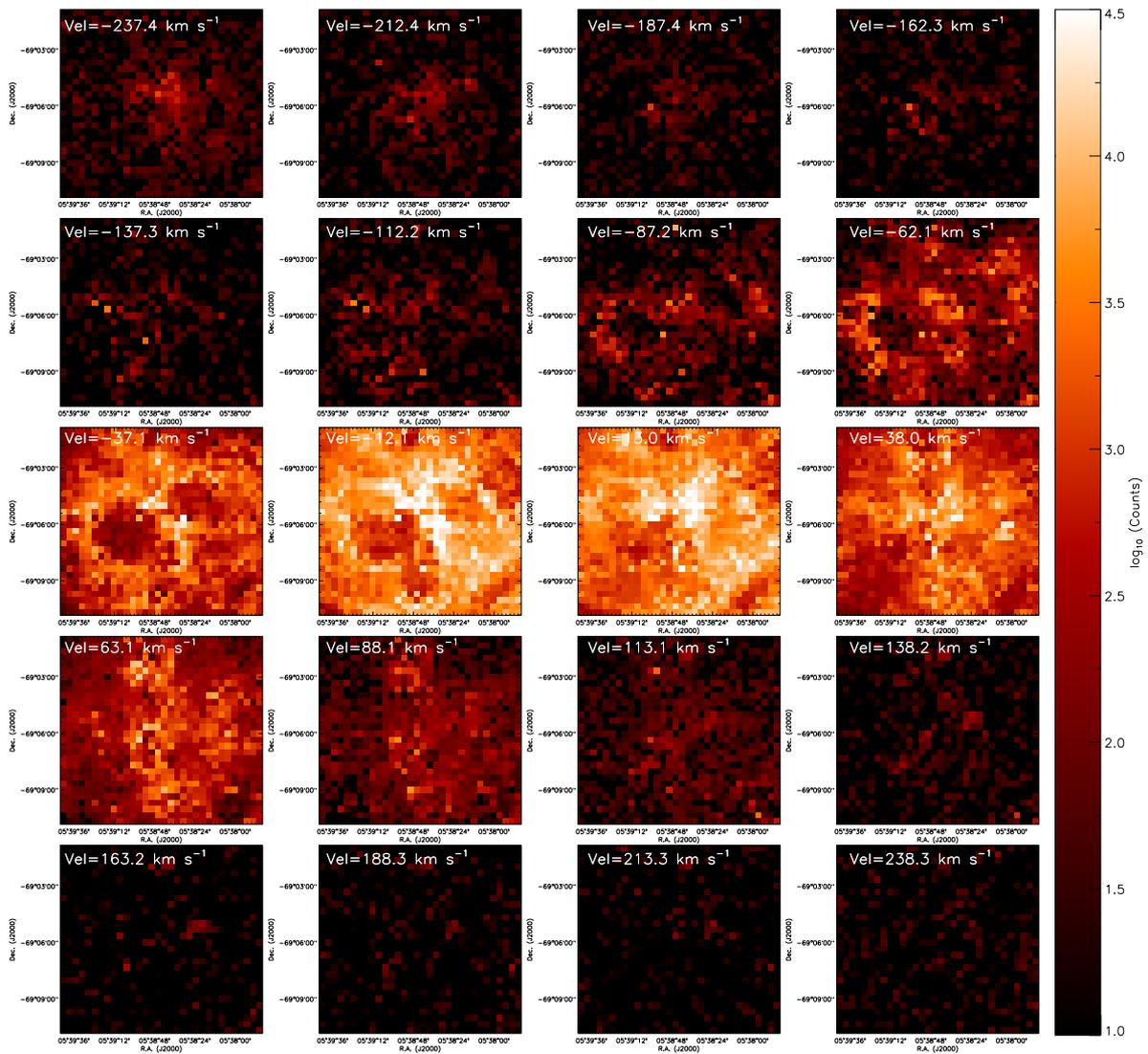}
\caption{Velocity-sliced view of the 30~Dor datacube
  centered on the [N{\sc ii}] $\lambda$6584 line. The velocity with
  respect to the center of the [N{\sc ii}] $\lambda$6584
  line is shown in the upper left-hand corner of each panel (negative and positive velocities imply radially approaching and receding velocities, respectively).
  In this figure, a logarithmic scale is used for the intensity of
  each map.}
\label{HI_frames_NII}
\end{figure*}

\section{Results}
\label{results}

\subsection{General view of the 30~Dor spectroscopic datacube}
\label{generalview}

To visualize the complex morphology of the ionized emission in 30~Dor,
in Fig.~\ref{plot_adhoc_cut_Halpha_image} we have plotted the
H$\alpha$ profile of each fiber from the regular grid (normalized by
the peak of the intensity in each case) over an H$\alpha$ image obtained with the EMMI instrument at the La Silla Observatory\footnote{This image was taken on December 2002, under the program 70.C-0435(A), using the filter Ha\#596, with a total exposure time of 1340 seconds.}.

In Fig.~\ref{plot_adhoc_cut_Halpha_image} it is possible to correlate
the high-resolution spectral information with features seen in the
image. The H$\alpha$ profiles in many instances are extremely complex.
For example, in the cavity located 1$'$ to the east of R136, the
H$\alpha$ profiles clearly display multiple components. In fact, the
fiber located at the position $\alpha$\,$=$\,05$^{\rm h}$38$^{\rm
  m}$53.21$^{\rm s}$ $\delta$\,$=$\,$-$69$^{\circ}$05$'$59.89$''$
(J2000) displays at least five components (see the last panel of
Fig.~\ref{profile_examples}).  In general, most of the multiple
H$\alpha$ profiles are located in the regions where expanding shells
have been previously identified (see Chu \& Kennicutt 1994). On the other hand, Fig. ~\ref{plot_adhoc_cut_Halpha_image} shows that the most intense H$\alpha$ emitting regions in 30~Dor (based on its EMMI image) display simple and narrow profiles. In several cases, these bright regions are associated with photodissociation regions (PDRs), which can halt some expanding structures, producing simple profiles. In addition, Fig.~\ref{plot_adhoc_cut_Halpha_image} shows that some low-intensity regions display simple profiles (for example, in the north-east region), which suggests that no expanding structures lie at that location. By inspecting Fig.~\ref{plot_adhoc_cut_Halpha_image} it is also possible
to identify low-/high-velocity components. For instance, at
$\alpha$\,$=$\,05$^{\rm h}$37$^{\rm m}$49.66$^{\rm s}$
$\delta$\,$=$\,$-$69$^{\circ}$02$'$39.91$''$ we find a low-intensity
component having an approaching velocity larger than 200 km s$^{-1}$
compared to the systemic value.

Given the kinematic richness of 30~Dor, we have derived a
velocity-sliced view of the datacube, centered on the H$\alpha$ and
[N{\sc ii}] $\lambda$6584 lines. In Fig.~\ref{HI_frames_Halpha} we show 20 frames of the datacube. The systemic (zero) velocity was defined by the Gaussian fit to the integrated
H$\alpha$ line (i.e. from Fig.~\ref{espectro_total_integrado}), in
which the line-center was 6568.65\,{\AA}, equivalent to $v_{\rm
  r}$\,$=$\,267.4\,km\,s$^{-1}$ (where the rest wavelength of the H$\alpha$ emission line was taken from Hirata \& Horaguchi 1995) and the velocity interval between each
frame was 25.1\,km\,s$^{-1}$. In this case, negative and positive velocities imply radially approaching and receding velocities, respectively.

By inspecting the first four frames of
Fig.~\ref{HI_frames_Halpha}, we can see several high-velocity components with
respect to the mean velocity of 30~Dor, especially in the neighborhood
of the Hodge\,301 cluster (the high velocity components in the first and second
frames). These components reach approaching velocities larger than
230\,km\,s$^{-1}$. By inspecting the datacube we found several
low-intensity components that present velocities even larger than
250\,km\,s$^{-1}$. For instance, Redman et al. (2003) found several
discrete high-speeds knots in 30~Dor, which have velocities of
$\pm$200\,km\,s$^{-1}$. In the last frames of
Fig.~\ref{HI_frames_Halpha} we also note several high-velocity components
-- some of these can be associated with the expanding
structure \#5 from Chu \& Kennicutt (1994, their Fig 1.c).

Another interesting feature that appears in 
Fig.~\ref{HI_frames_Halpha} is the detection of shell-like structures which
are visible across all the observed field of 30~Dor (see section \S
\ref{expandingstructures}). These structures appear as small voids
that increase in size as we move from the blue to the red side of the
H$\alpha$ line. It is interesting to note that bluewards of the
systemic H$\alpha$ emission the shells are clearly defined, while they
are not seen in the frames redwards of the central velocity. Internal
extinction produced by dust can produce this signature. Also,
molecular clouds located at these positions could hide the red side of
the H$\alpha$ emission.

\begin{figure}[h!]
\centering
\includegraphics[width=\columnwidth]{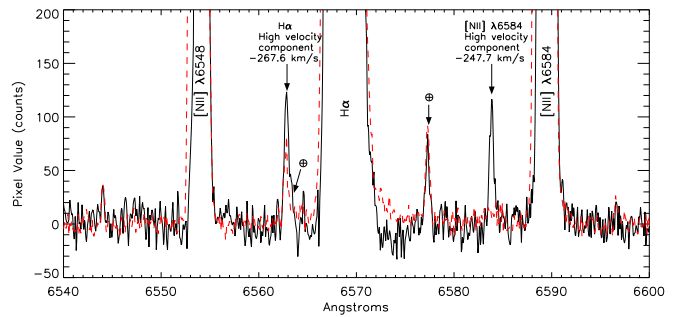}
\caption{An example of a high-velocity component seen in the [N{\sc ii}] $\lambda$6584 line. The black solid and red dashed lines correspond to the spectra of two contiguous fibers on the sky. The black solid spectrum shows a [N{\sc ii}] high velocity component which is intense in comparison with its H$\alpha$ counterpart. This effect is not detected in the red dashed spectrum.}
\label{high_vel}
\end{figure}

In Fig.~\ref{HI_frames_NII}, we show a similar
velocity-sliced view, but now centered on the [N{\sc ii}]
$\lambda$6584 emission line. As in the previous case, the systemic (zero) velocity was derived from the Gaussian fit to the integrated [N{\sc ii}] $\lambda$6584 line, which gives a radial velocity of $v_{\rm  r}$\,$=$\,266.1\,km\,s$^{-1}$ (6589.29\,{\AA}). As to be expected from the reduced
intensity of the [N{\sc ii}] line compared to H$\alpha$, most of the
emission is detected at lower expanding velocities than seen in the
H$\alpha$ map. While there is a clear correlation between the
morphology of the shell-like structures detected in both lines, 
inspection of Fig.~\ref{HI_frames_Halpha} and Fig.~\ref{HI_frames_NII}
shows that several H$\alpha$ and [N{\sc ii}] $\lambda$6584 high-velocity
components do not correlate spatially, which suggests that different
physical processes are exciting the gas at these locations. For example, in Fig. \ref{high_vel} we show a high-velocity component in the [N{\sc ii}] $\lambda$6584 line (black solid line) which presents a strong intensity with respect to their H$\alpha$ counterpart ($\alpha$\,$=$\,05$^{\rm h}$38$^{\rm m}$41.97$^{\rm s}$
$\delta$\,$=$\,$-$69$^{\circ}$04$'$40.04$''$). In the same figure, we see the spectrum of the next observed fiber (red dashed line), which present a high-velocity component in H$\alpha$ but with no [N{\sc ii}] $\lambda$6584 emission ($\alpha$\,$=$\,05$^{\rm h}$38$^{\rm m}$41.96$^{\rm s}$
$\delta$\,$=$\,$-$69$^{\circ}$05$'$00.03$''$). We note that in this comparison we are using two adjacent observed fibers. These
points all help to illustrate the capabilities of these data in
disentangling the complex structures of 30~Dor.

\subsection{Complexity of the H$\alpha$ profiles in 30 Doradus}
\label{complexityprofiles}

\subsubsection{From single to multiple profiles}
\label{sincesinlge}

\begin{figure*}
\begin{center}
\includegraphics[width=0.44\textwidth]{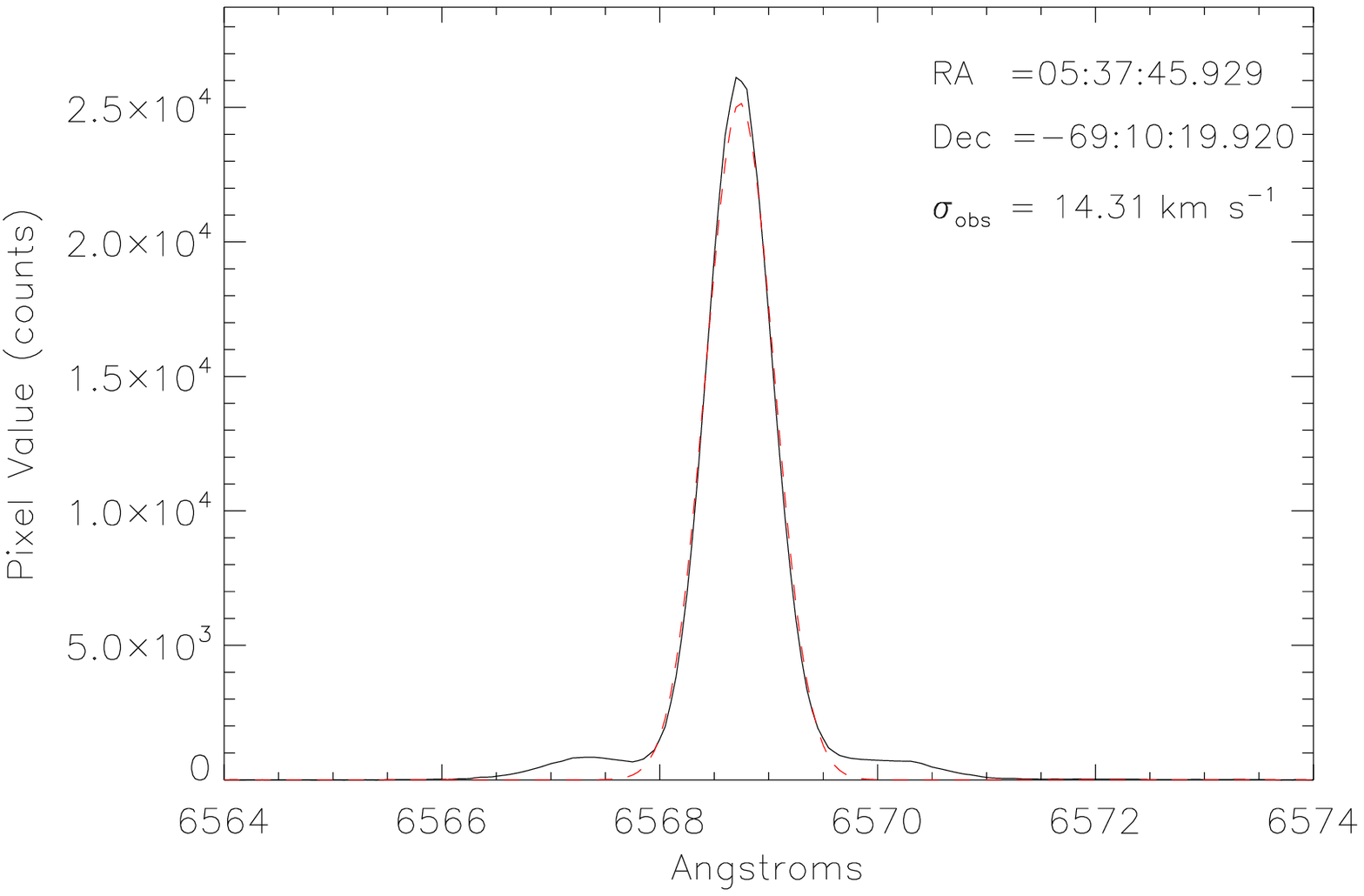}
\includegraphics[width=0.44\textwidth]{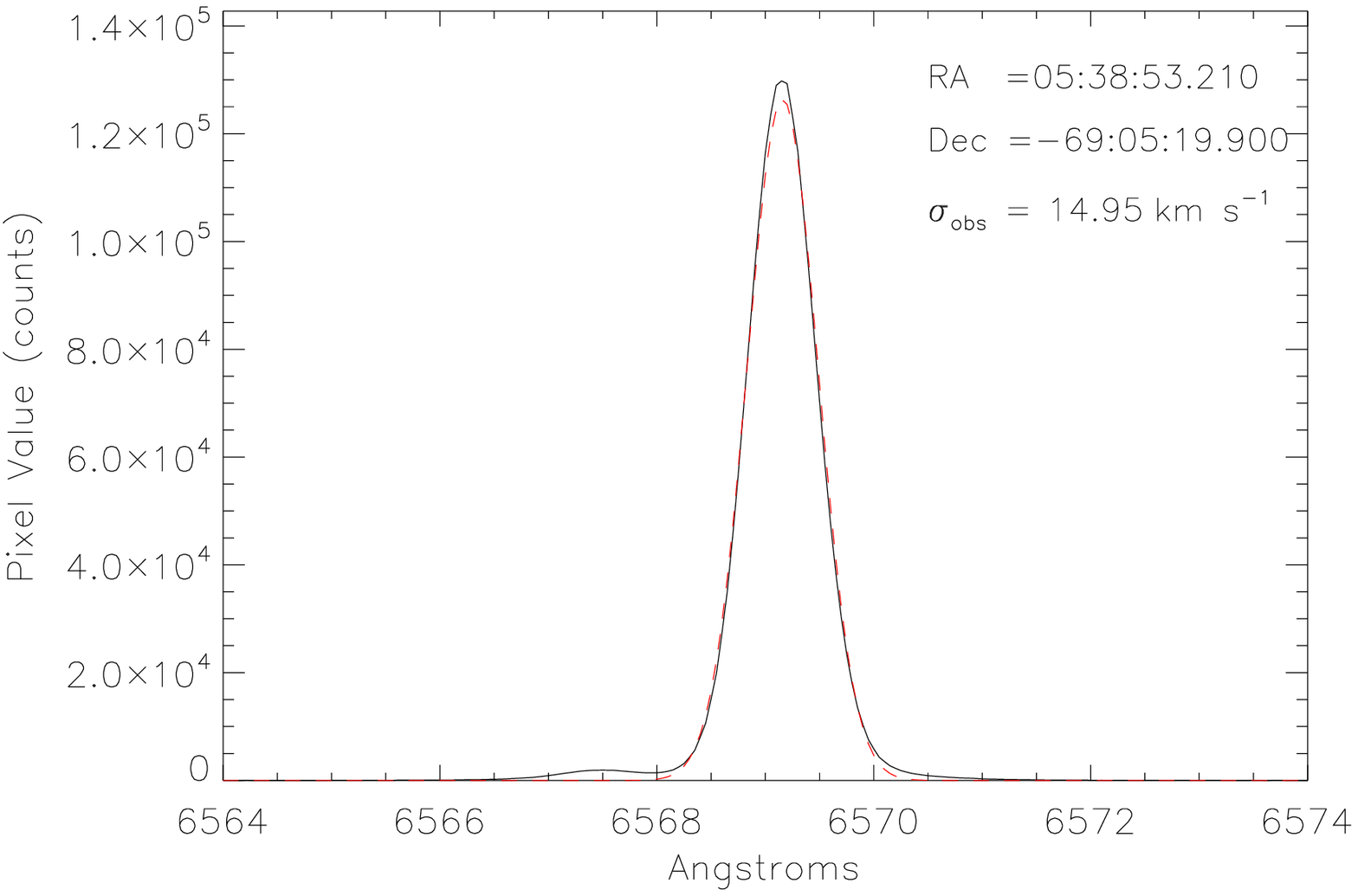}\\
\includegraphics[width=0.44\textwidth]{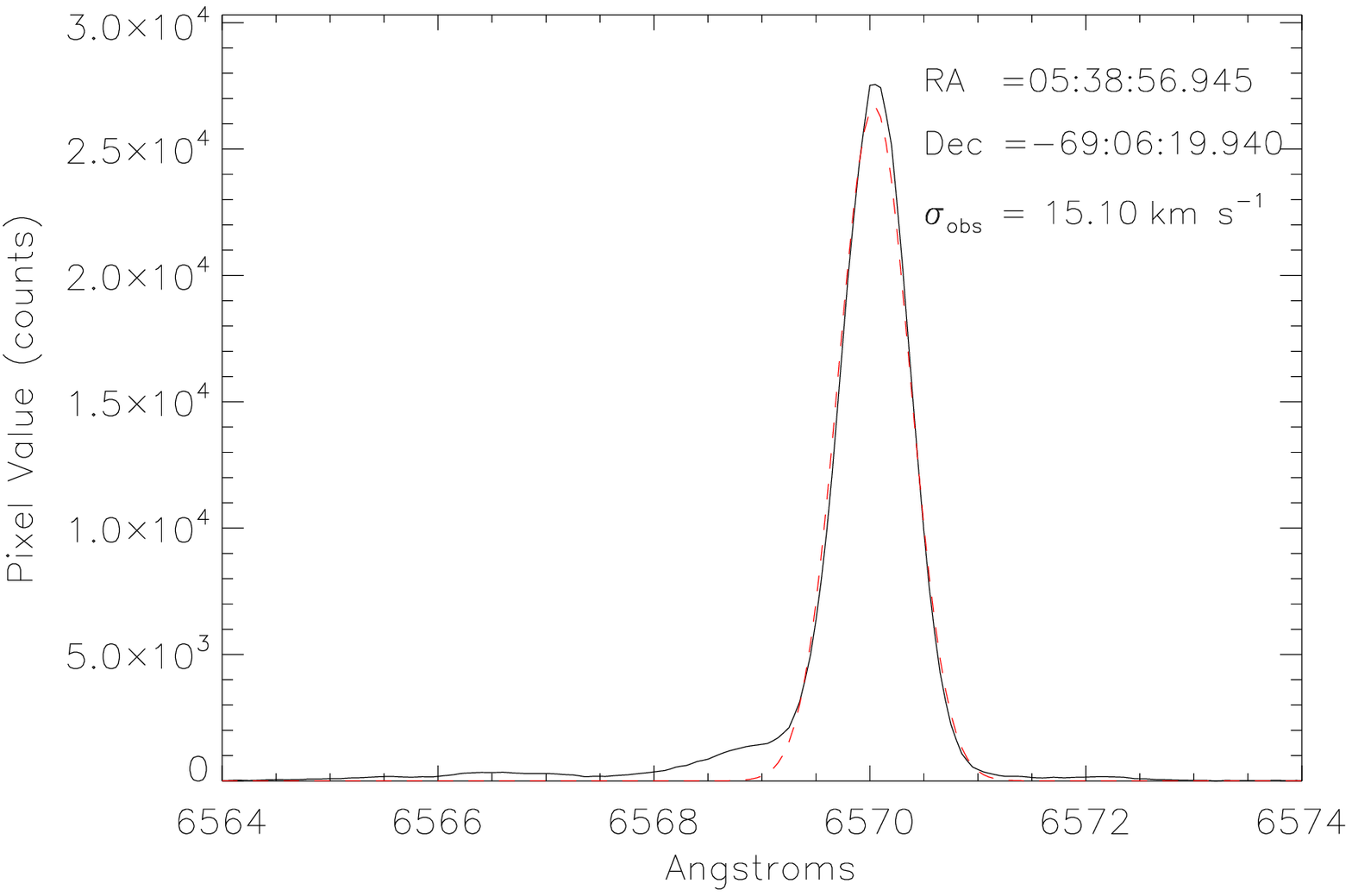}
\includegraphics[width=0.44\textwidth]{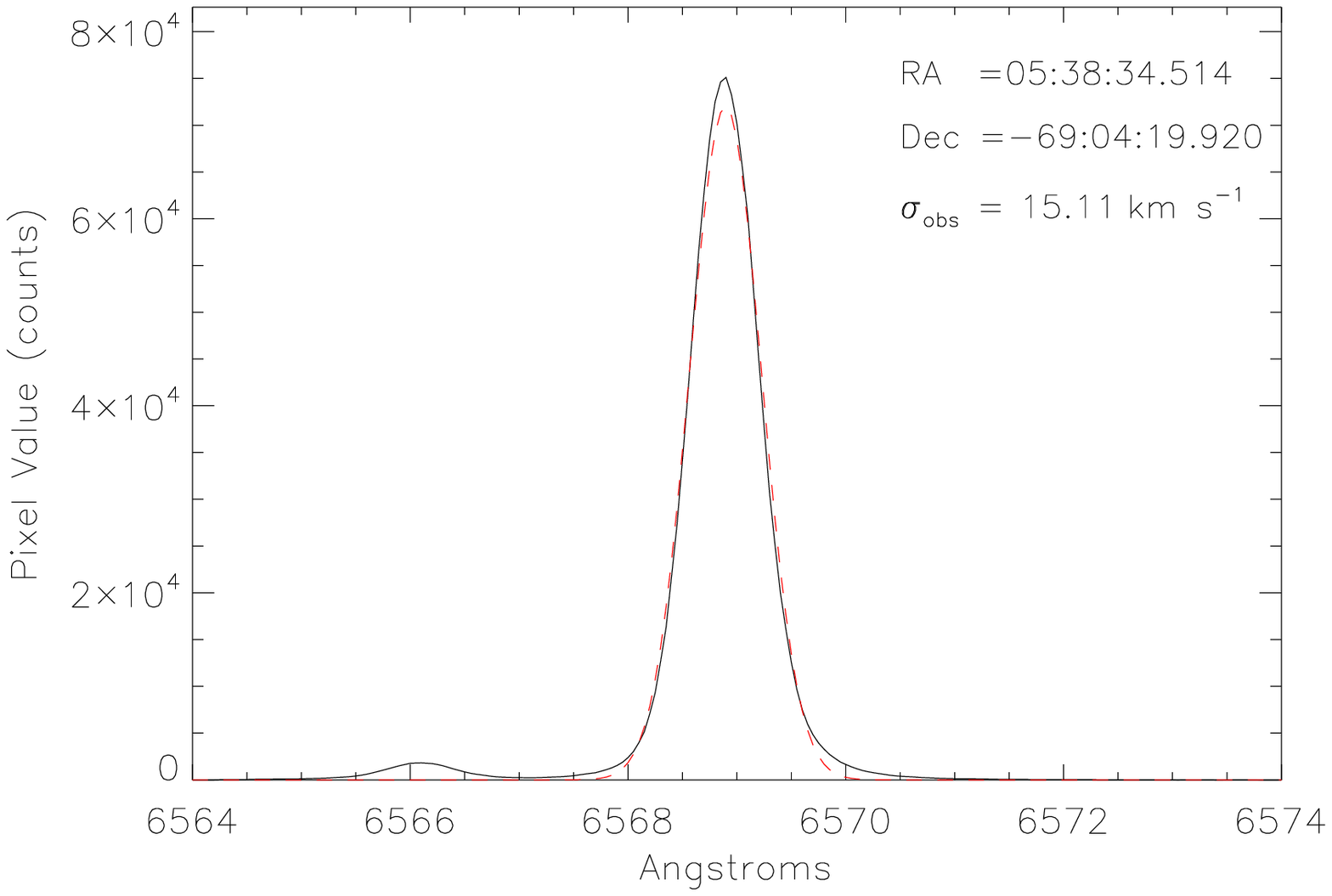}\\
\includegraphics[width=0.44\textwidth]{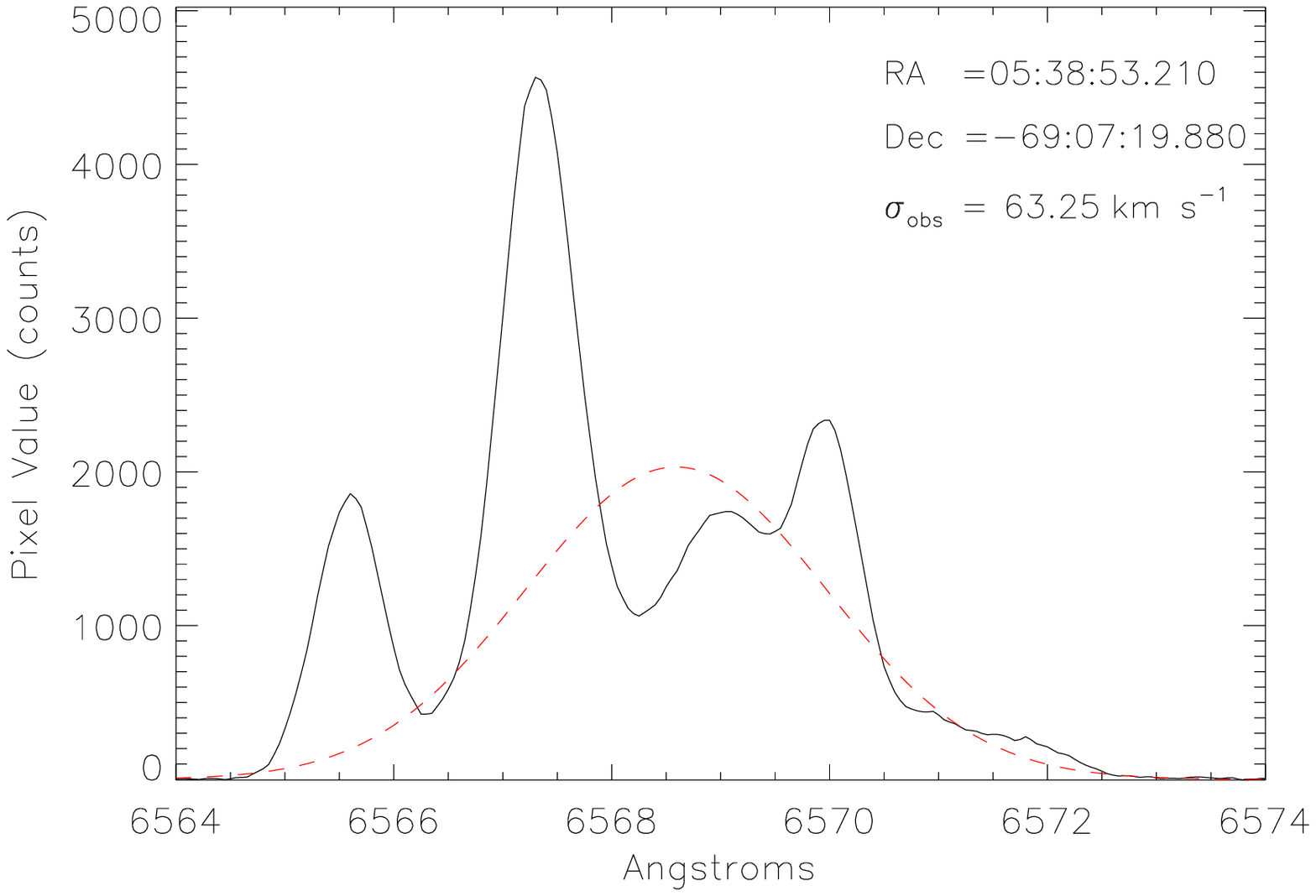}
\includegraphics[width=0.44\textwidth]{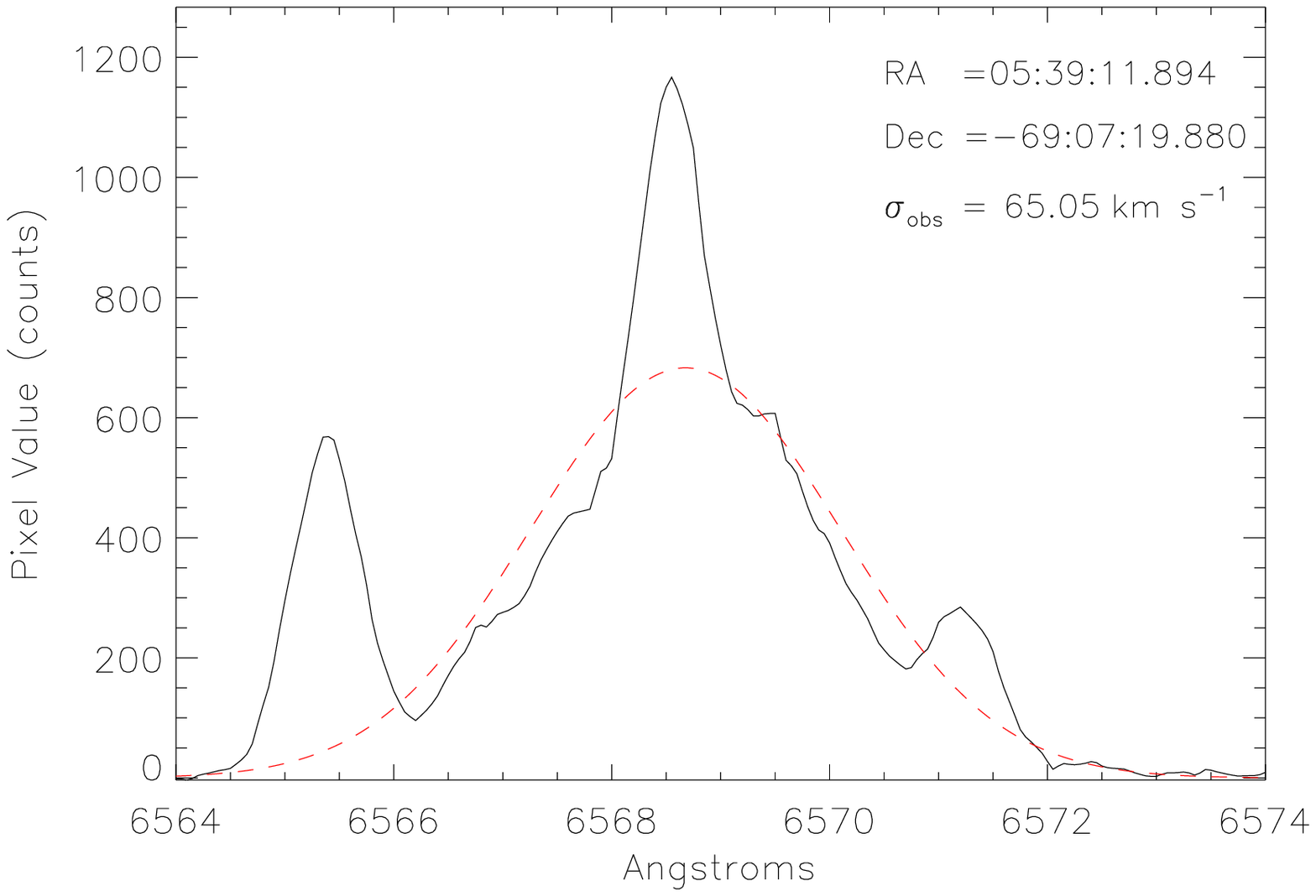}\\
\includegraphics[width=0.44\textwidth]{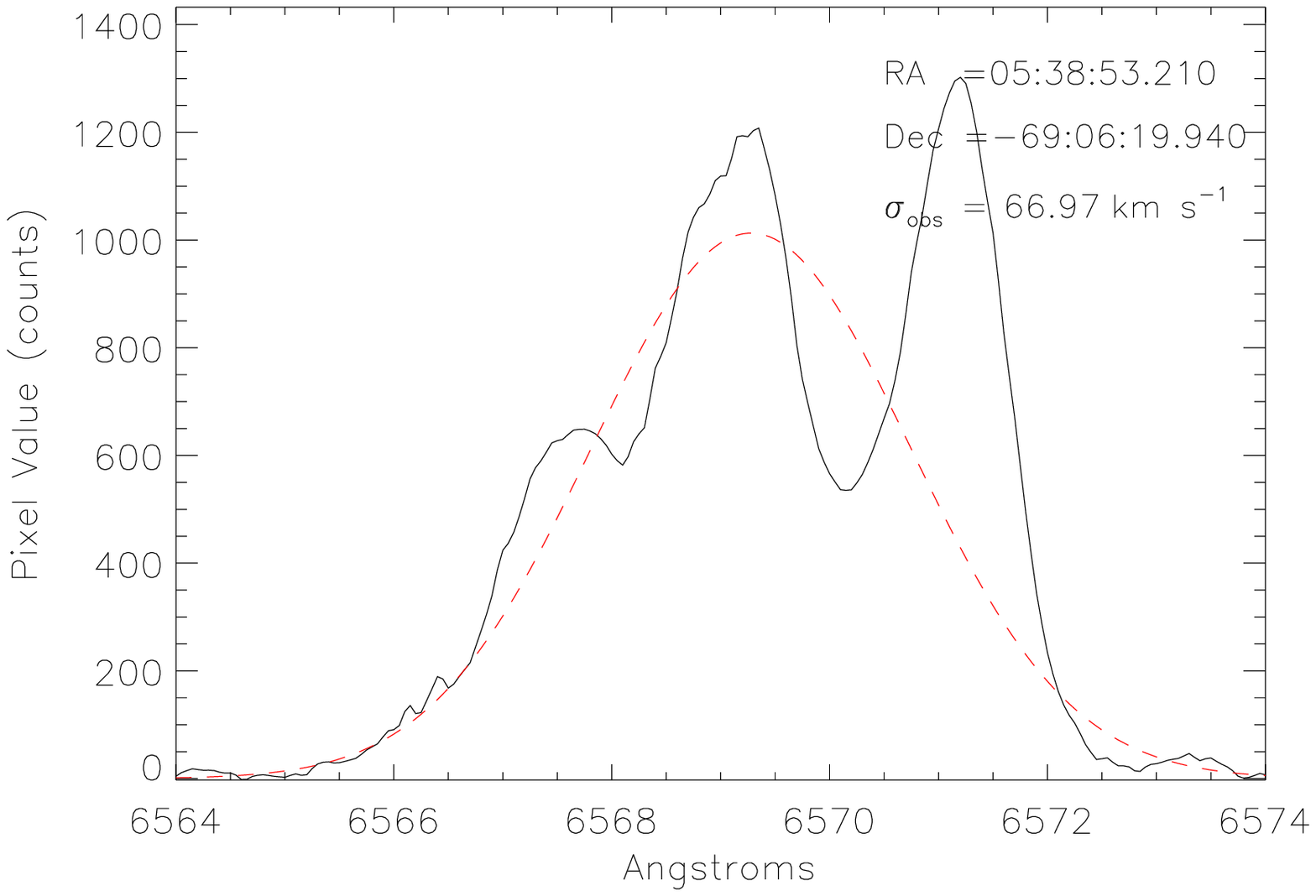}
\includegraphics[width=0.44\textwidth]{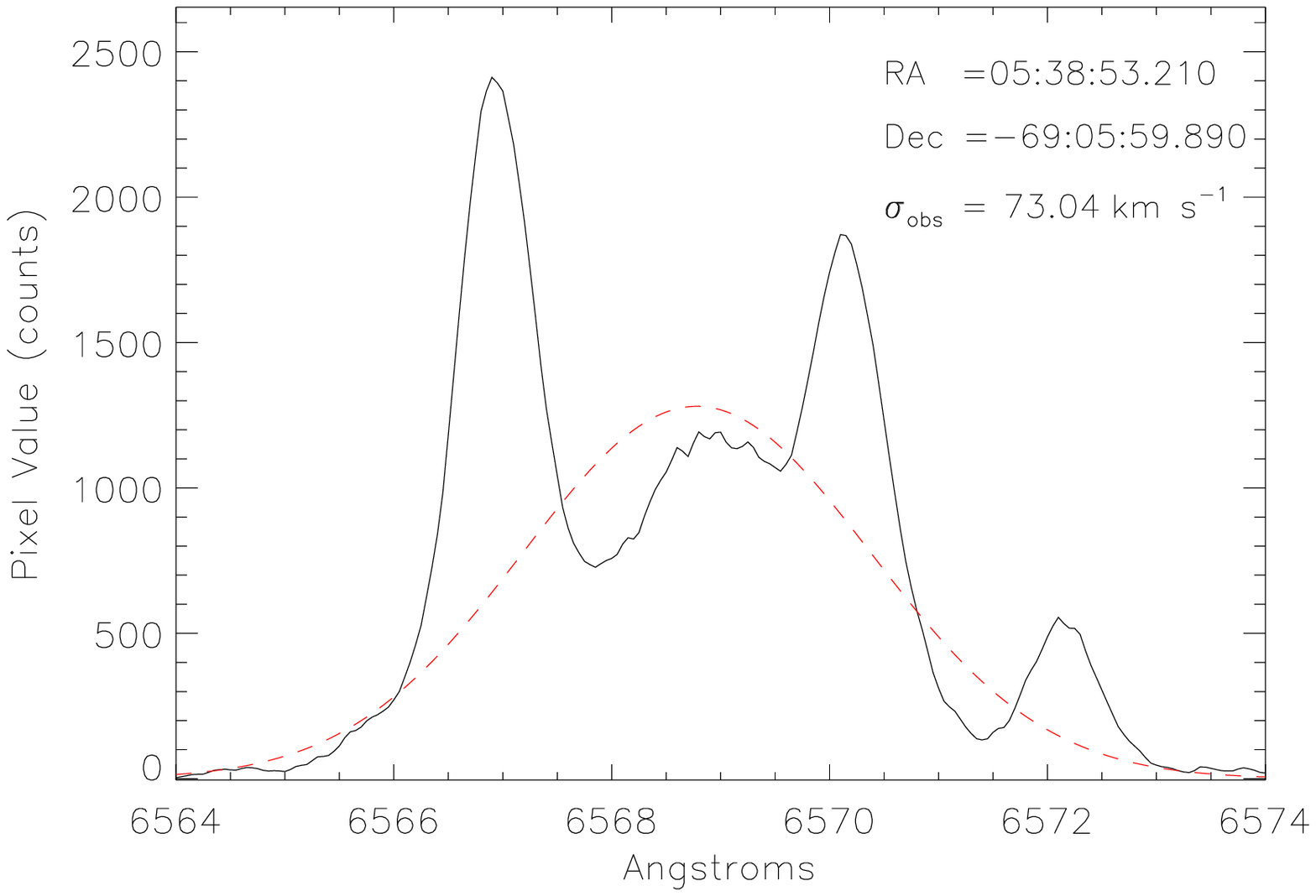}
  \caption{Examples of the different H$\alpha$ morphologies detected
    in 30~Dor. In the first four panels we show the narrowest
    H$\alpha$ profiles found in this study. In contrast, the last four panels show the broadest velocities arising from single-profile fits (which are clearly a consequence of multiple components in the observed lines). In each panel we note the coordinates of the observation and $\sigma_{\rm obs}$ of the single-profile Gaussian fit (indicated by the overplotted red dashed lines).}
\label{profile_examples}
\end{center}
\end{figure*}

The wide variety of H$\alpha$ profiles in the data are
illustrated by the examples in Fig.~\ref{profile_examples},
which show the four narrowest and broadest profiles (upper and lower
panels, respectively). All these profiles correspond to those observed and not to the averaged profiles described in \S~\ref{regulargrid}. A single Gaussian was fitted to each observed
profile (as described in \S~\ref{analysis}), with the coordinates of
the fiber position and $\sigma_{obs}$ derived from the fitting process
shown in the upper right of each panel. In this analysis, the center of the Gaussian fit gives the overall systemic radial velocity at the position of the fiber, while the width of this fit gives the velocity dispersion of the ionized gas, which must be corrected by the instrumental and thermal widths.

The narrowest H$\alpha$ profile has a
$\sigma_{obs}$\,$=$\,14.3\,km\,s$^{-1}$ (upper left-hand panel in
Fig.  \ref{profile_examples}), which corresponds to
$\sigma$\,$=$\,7.8\,km\,s$^{-1}$ once corrected for $\sigma_{th}$ and
$\sigma_{in}$. This profile is located in the south-west of the
nebula, which is ionized by the LH99 stellar association (Lucke \&
Hodge 1970) and is also the location of the SN remnant 30~Dor~B
(Danziger et al. 1981).  Two low-intensity components can also be seen
in the profile, and these could be associated with an expanding structure.
Nonetheless, the main structure of this profile is symmetric and can
be well-fitted by a single Gaussian (red dashed line in
Fig.~\ref{profile_examples}). The other three narrow H$\alpha$
profiles shown in Fig.  \ref{profile_examples} display a few
low-intensity components, but the main part of the profiles are also
well-fitted by a single Gaussian.

In contrast, the broad profiles shown in Fig.~\ref{profile_examples}
clearly display multiple strong components. As expected, these
profiles can not be fit by a single Gaussian and their measured widths
are the result of a coarse fit that takes into account all the
components. All of these broad, multiple profiles lie in shell-like
structures. It is interesting to note that the third broad profile
shown in Fig.~\ref{profile_examples} lies just 20$''$ (equivalent to
$\sim$5\,pc) from the narrow profile located at
$\alpha$\,$=$\,05$^{\rm h}$38$^{\rm m}$56.945$^{\rm s}$
$\delta$\,$=$\,$-$69$^{\circ}$06$'$19.940$''$.  This shows the
extremely complex structure of 30~Dor even at small spatial scales.
 
\begin{figure*}
\centering
\includegraphics[width=0.48\textwidth]{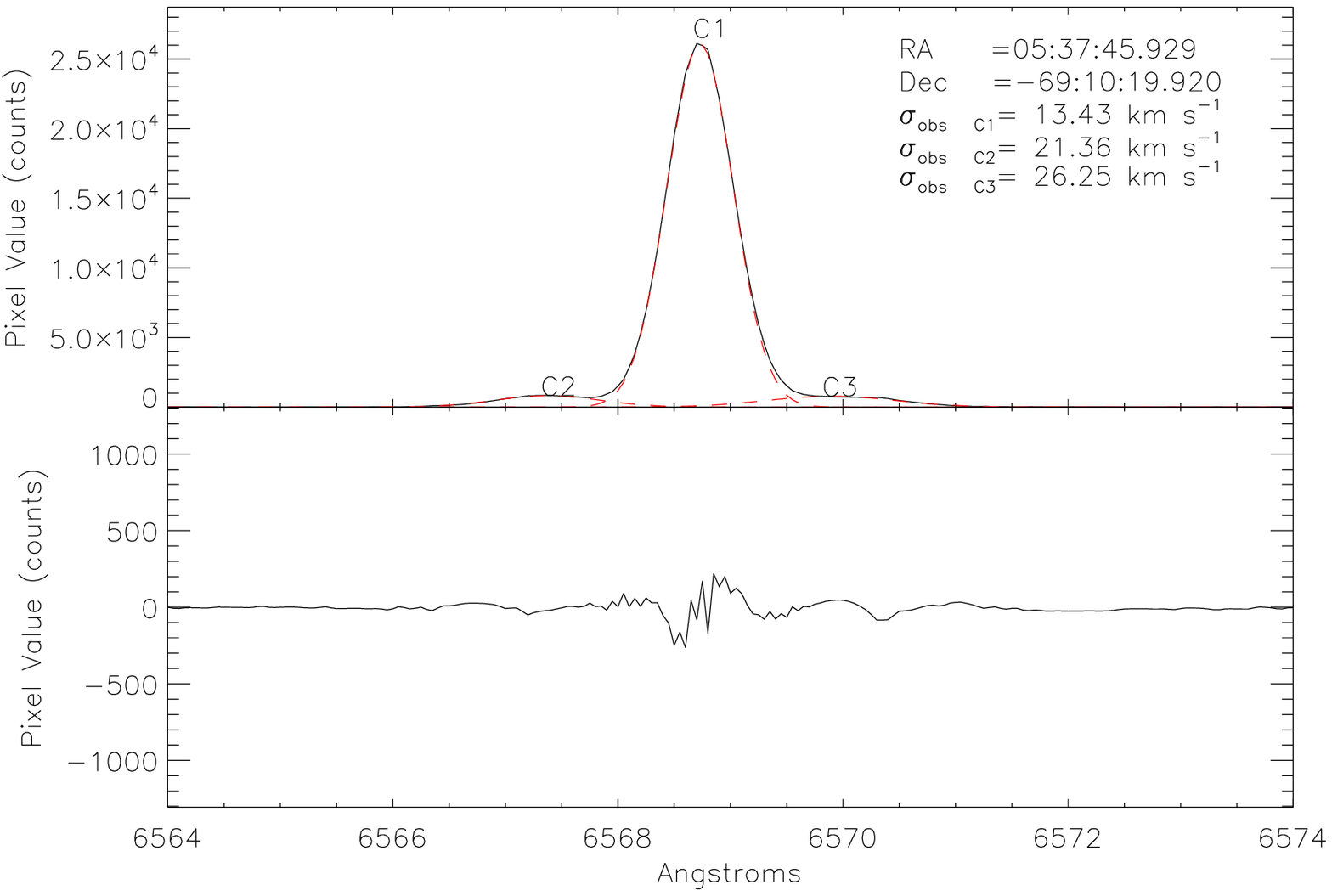}
\includegraphics[width=0.48\textwidth]{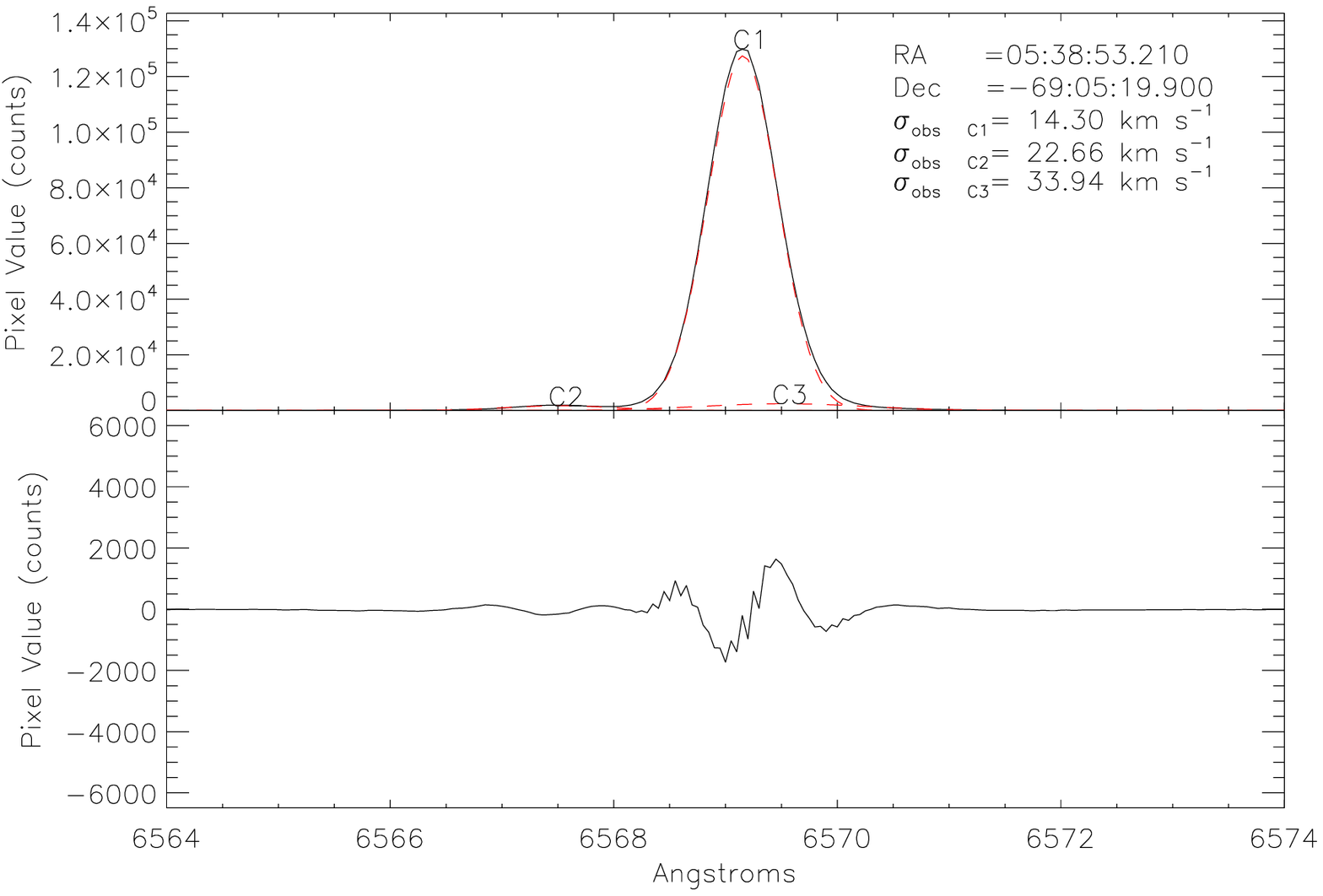}\\
\includegraphics[width=0.48\textwidth]{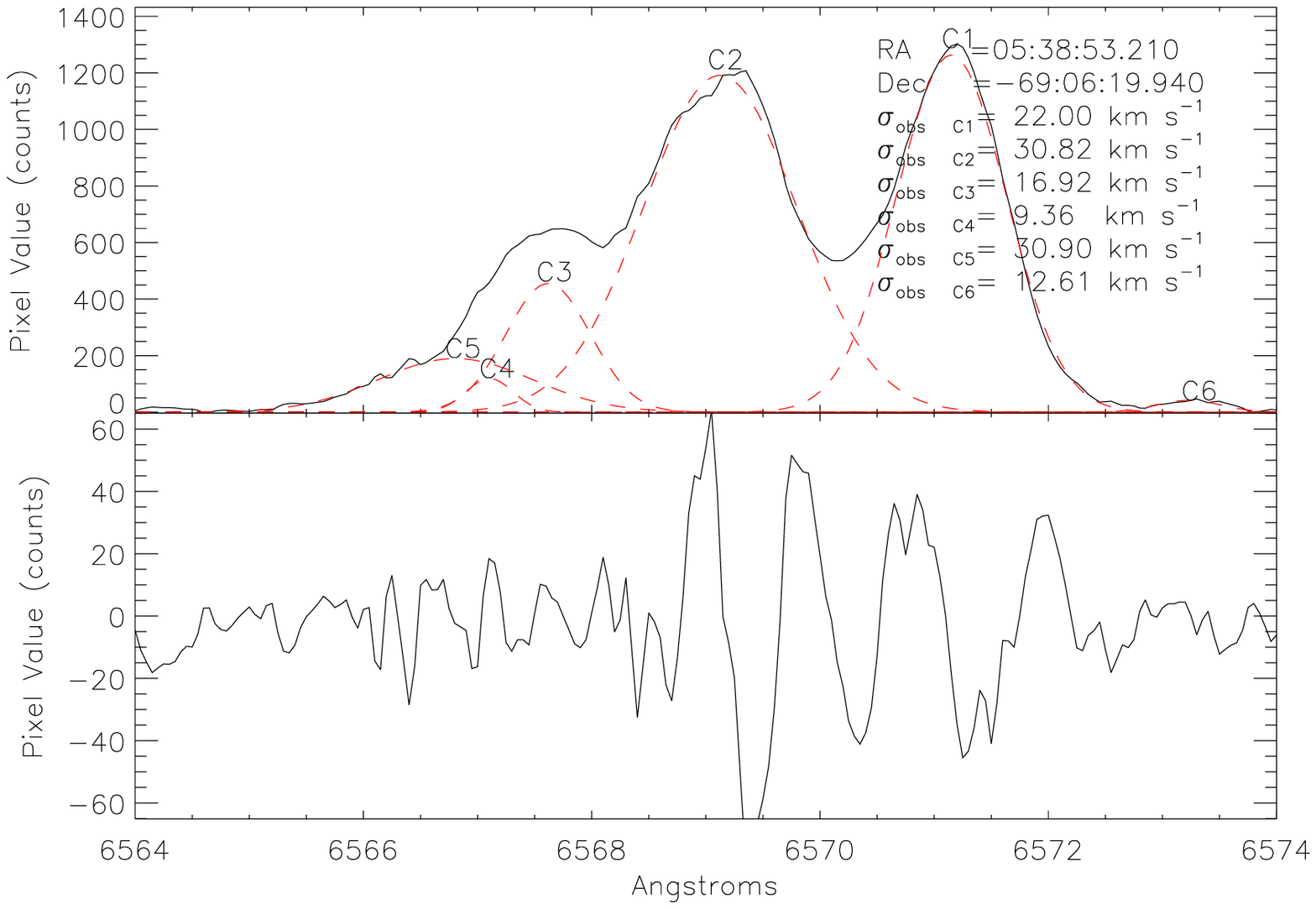}
\includegraphics[width=0.48\textwidth]{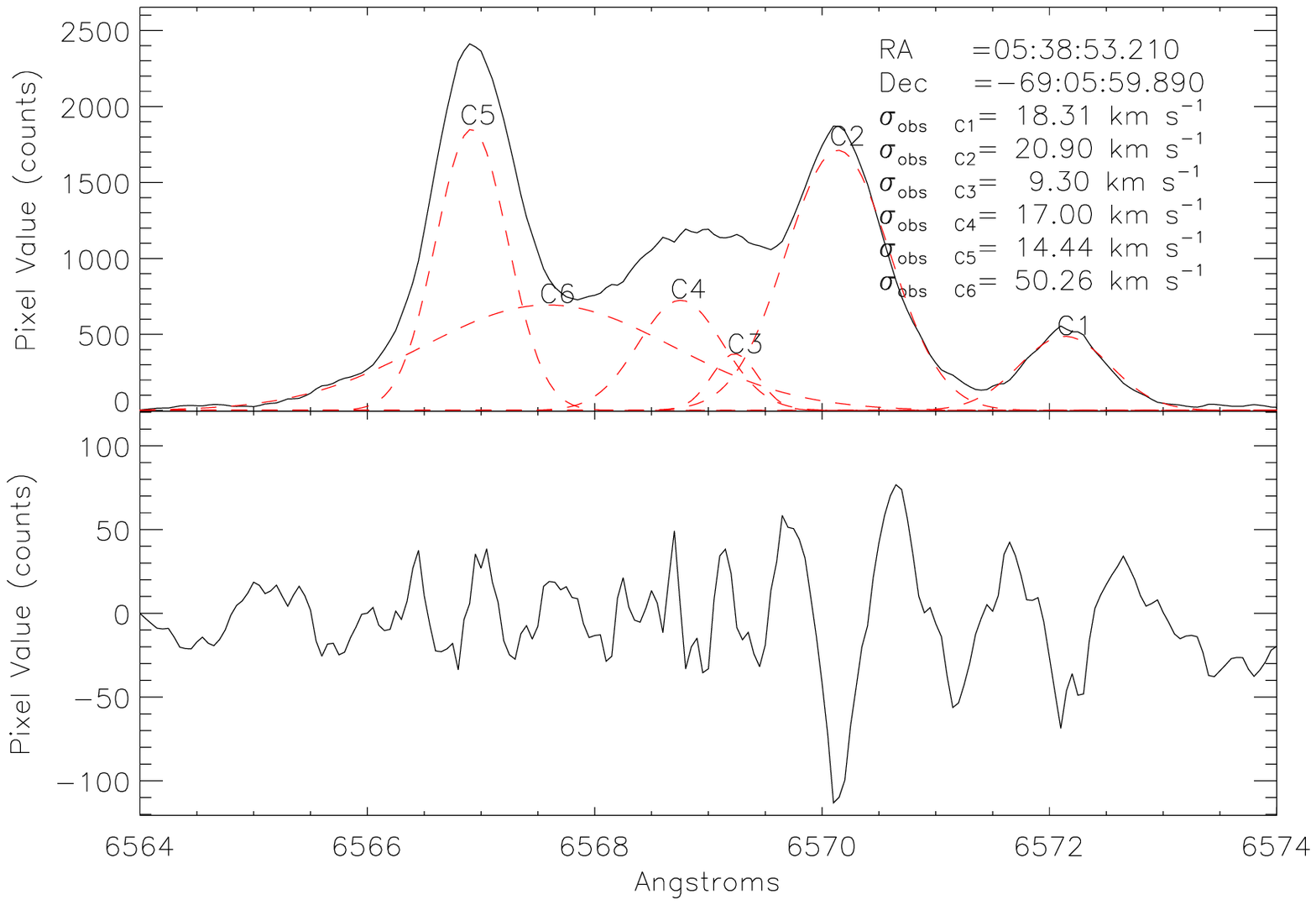}
\caption{Multiple Gaussian fits to the H$\alpha$ profiles in 30~Dor with the narrowest and broadest single-component velocity dispersions  (upper and lower panels, respectively).
  The black (solid) and red (dashed) lines represent the observed and
  fitted profiles, respectively; each component is numbered.
  Residuals of the fits are shown below each fit, which are relatively
  minor when compared with the intensity of the profiles. In each
  panel we list the position of the spectra and the $\sigma$ of the
  different fitted components.}
\label{profiles_pan}
\end{figure*}

\subsubsection{Multiple Gaussian fits: A few examples}
\label{multiplegaussians}

To determine the line-broadening mechanism in 30~Dor, Melnick et al.
(1999) fitted multiple Gaussian components to the observed H$\alpha$
profiles of different regions. They found that a broad, low-intensity
component was necessary to reproduce the wings of all the observed
H$\alpha$ profiles.  As noted earlier, we have performed multiple
Gaussian fits to a subset of profiles using the {\sc pan} package.  In
particular, we have examined the two narrowest and broadest H$\alpha$
profiles from single-component fits on the FLAMES data (see Fig.~\ref{profile_examples}), with
results of these fits shown in Fig.~\ref{profiles_pan}. In each panel
of the figure the red (dashed) lines indicate the different Gaussian
components used in the fits; the coordinates of the fiber position and
the $\sigma_{obs}$ of each fitted component are given in the upper
right of the panels. In Fig.~\ref{profiles_pan} we also plot the
residuals of the fits, i.e.  the difference between the fitted and the
observed profile, in which the intensity axis was limited to $\pm$5\%
of the peak of the observed profile.

For the narrowest H$\alpha$ profile in the data (upper left in
Fig.~\ref{profiles_pan}), we have fitted three Gaussian components.
The two low-intensity components detected at this position present
widths smaller than $\sigma_{obs}$\,$\sim$\,26\,km\,$s^{-1}$. The
strongest component at this position has a corrected width of
$\sigma$\,$=$\,6.1\,km\,s$^{-1}$, i.e., lower than the value obtained
by fitting just one Gaussian ($\sigma$\,$=$\,7.8\,km\,s$^{-1}$, from
Fig.~\ref{profile_examples}). By inspecting the residual of these
fits, no low-intensity broad component is necessary to explain the
observed profile (the residuals are negligible compared with the
observed emission). This is of wider interest as this H$\alpha$
profile is located in 30~Dor~B, associated with a SN remnant (where we can expect a complex kinematic). In the case of the second-narrowest profile, we have fitted three Gaussian components. The central, most intense feature is fit with a narrow component with a corrected width of 7.8\,km\,s$^{-1}$, together with two broader, low-intensity components with corrected widths of 19.2\,km\,s$^{-1}$ and 31.8\,km\,s$^{-1}$.

The scenario is quite different when multiple Gaussians are fitted to
the broadest and more complex H$\alpha$ profiles. In the lower
left-hand panel of Fig.~\ref{profiles_pan}, we show that at least six
components are necessary to reproduce the observed emission line at
that position. The broadest component has a corrected width of
$\sigma$\,$=$\,28.5\,km\,s$^{-1}$.  As in the previous cases, the
residual is negligible compared with the intensity of the profiles.
Finally, in the lower right-hand panel of Fig.~\ref{profiles_pan} we
show another spectrum which requires six components to fit the
observations. In this case, the broadest component has a corrected
width of $\sigma$\,$=$\,48.8\,km\,s$^{-1}$. It is interesting to note that four of the six broad components used in the Gaussian fits shown in the lower panel of Fig.~\ref{profiles_pan} display supersonic widths (once corrected by $\sigma_{th}$ and $\sigma_{ins}$). Despite other Gaussian components can be fitted in these wide profiles, our results suggest the presence of a highly turbulent gas at these locations.

We note that Melnick et al. (1999) found that most of the H$\alpha$
profiles in their analysis required the presence of a low-intensity
broad component with $\sigma$\,$=$\,45\,km\,s$^{-1}$, Moreover, this
broad component was required to fit their integrated H$\alpha$ profile
of 30~Dor, centered at the radial velocity of this GHR. This component
was not identified in the two narrowest profiles observed by FLAMES,
nor in the second broadest profile (shown in Fig.~\ref{profiles_pan}).
In fact, the broad components that we have identified are not at the
systemic radial velocity of the 30~Dor gas; the origin of the broad
component in 30~Dor, if real, is still a mystery.

\begin{figure}[t!]
\begin{center}
\includegraphics[width=0.9\columnwidth]{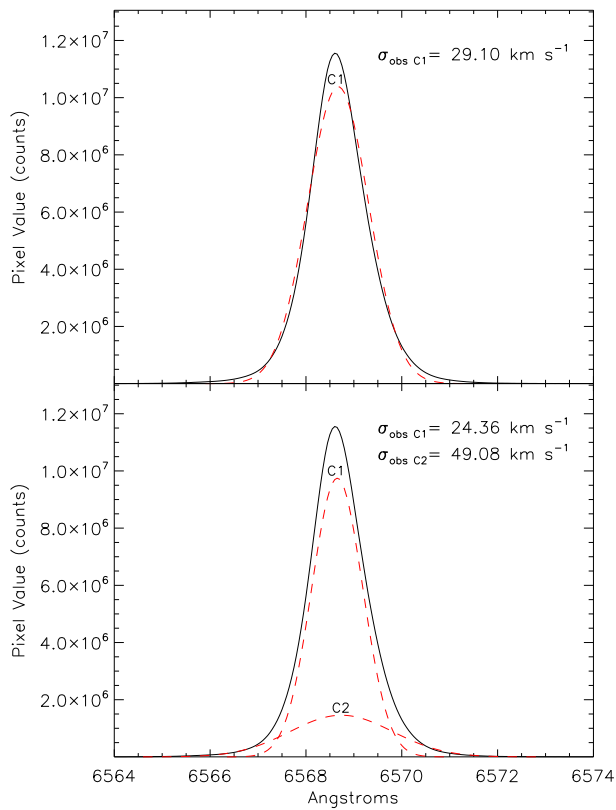}
  \caption{Single and two-component Gaussian fits to the integrated
    H$\alpha$ profile (upper and lower panels, respectively). The
    fitted components are shown by the red (dashed) lines and their
    widths are indicated in the upper right of each panel.}
\label{espectro_total}
\end{center}
\end{figure}

\subsubsection{The integrated H$\alpha$ profile of 30 Doradus}
\label{integratedprofile}

Fits to the integrated H$\alpha$ profile (from
Fig.~\ref{espectro_total_integrado}) are shown in
Fig.~\ref{espectro_total}.  For a single-component fit we found a
corrected width of $\sigma$\,$=$\,26.5\,km\,s$^{-1}$ ($\sigma_{\rm obs}$\,$=$\,29.1\,km\,s$^{-1}$). Note from the
figure that the blue and red wings of the integrated profile can not be fitted by one component -- a second, broader component is required, as
shown in the lower panel of Fig.~\ref{espectro_total}.  By using
narrow, high-intensity and broad, low-intensity components, we obtain
a much better fit to the observed H$\alpha$ profile of 30~Dor. In
this case, the narrow and the broad components have corrected widths
of $\sigma$\,$=$\,21.2\,km\,s$^{-1}$ ($\sigma_{\rm obs}$\,$=$\,24.4\,km\,s$^{-1}$) and
$\sigma$\,$=$\,47.6\,km\,s$^{-1}$ ($\sigma_{\rm obs}$\,$=$\,49.1\,km\,s$^{-1}$), respectively. Our result for the narrow component is in good agreement with the corrected width from Melnick et al. (1999, $\sigma$\,$=$\,22\,km\,s$^{-1}$) from the data of Chu \& Kennicutt (1994); our broader component is marginally wider than the $\sigma$\,$=$\,44\,km\,s$^{-1}$ from Melnick et al. By comparing the bottom panel of Fig.~\ref{espectro_total} with the
Fig.~4 from Melnick et al., we note that our broad component is less
intense (with respect to the height of the profile) than their broad
fit, perhaps due to the different spatial extents that the two studies
cover.

\begin{figure}[t!]
\centering
\includegraphics[width=0.49\textwidth]{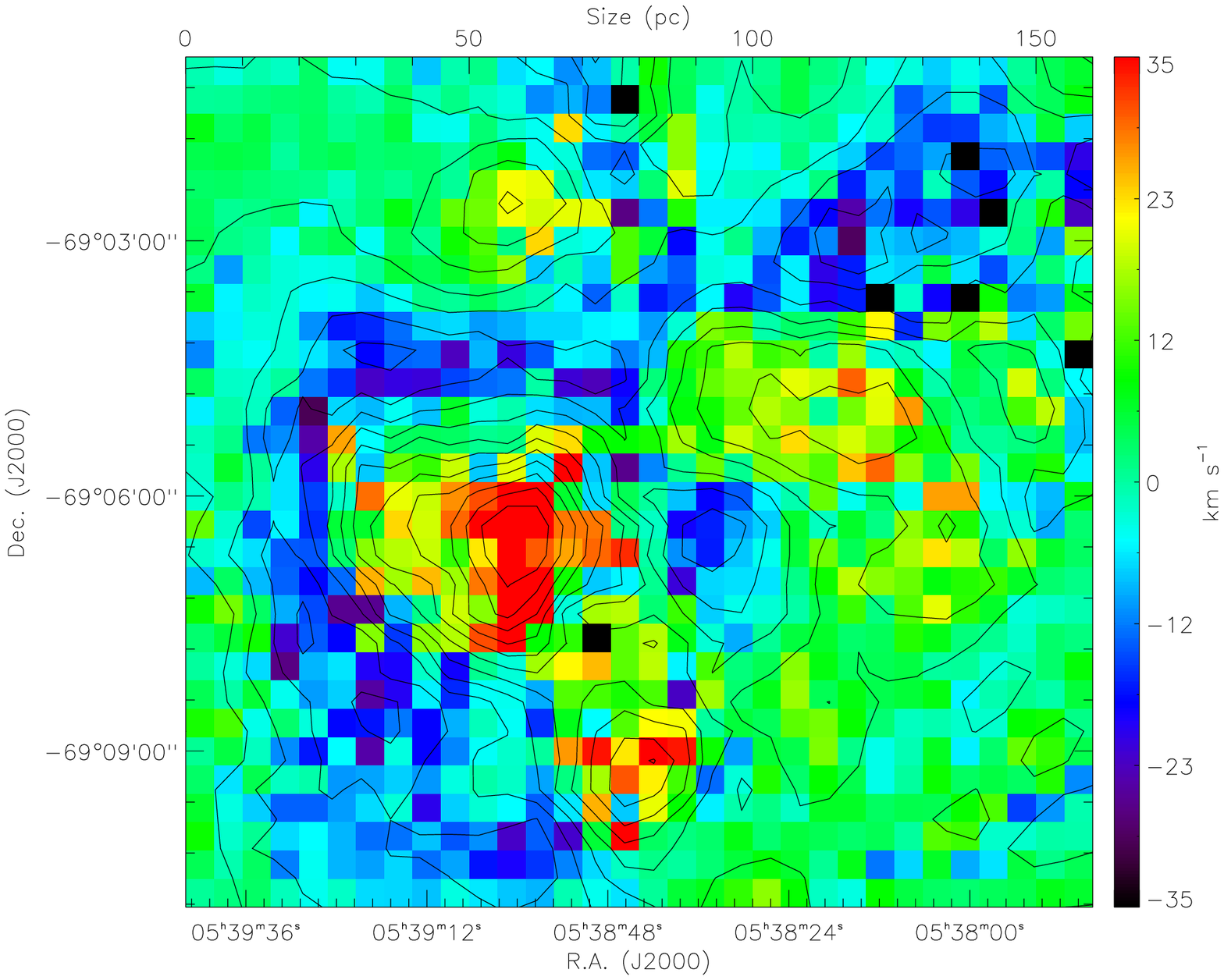}\\
\includegraphics[width=0.49\textwidth]{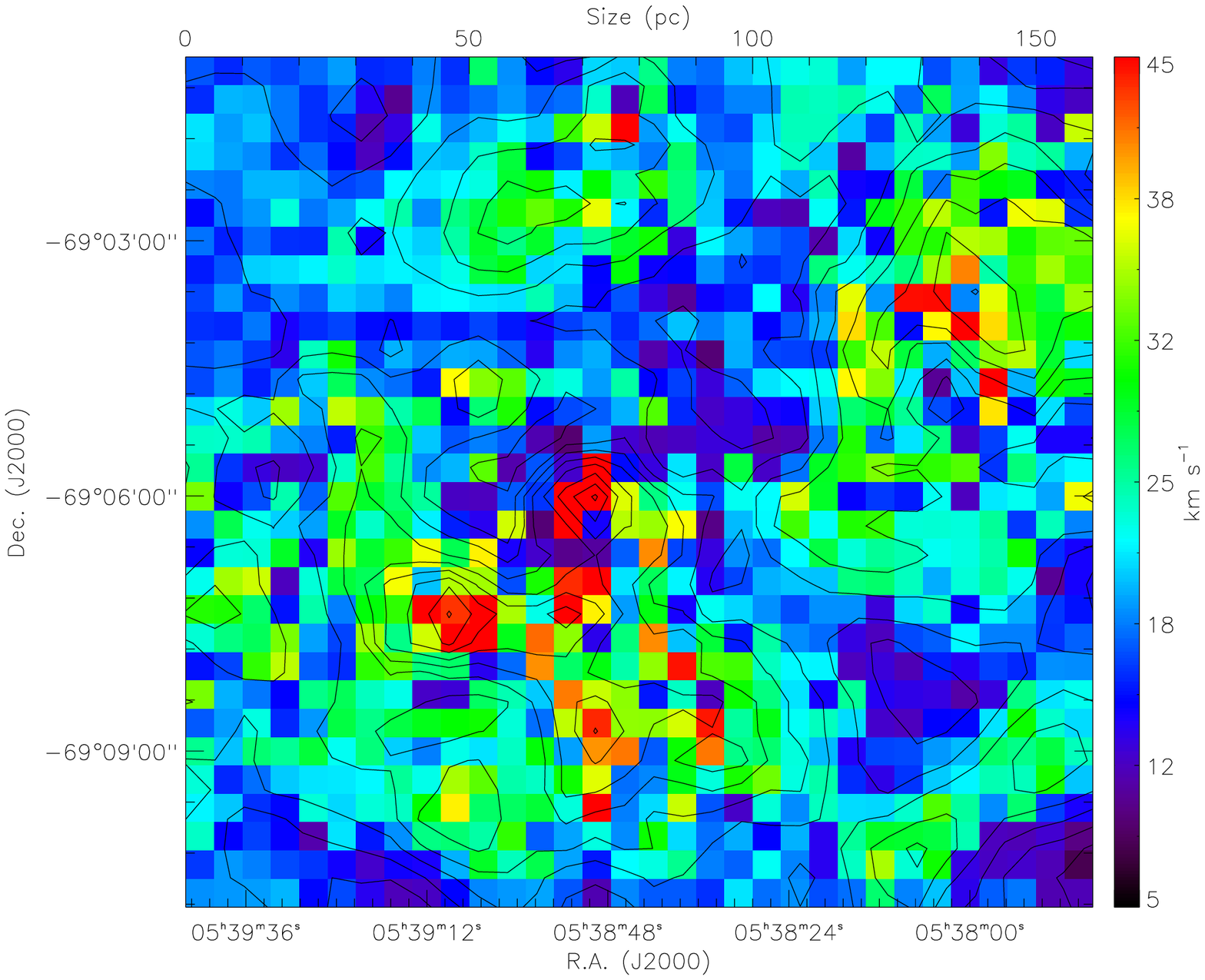}
\caption{Radial line-of-sight systemic velocity map (upper panel) and velocity dispersion map (lower panel) of 30~Doradus from analysis of the H$\alpha$ emission in the
  FLAMES spectra. North at the top and east to the left. The radial velocity map was corrected by the systemic velocity of 30~Dor.}
\label{velocity_map_halpha}
\end{figure}

\subsection{The velocity map of 30 Doradus}
\label{velocityfield}

As discussed in \S~3, we have fitted a single-component Gaussian to
each H$\alpha$ profile. We have used the center and $\sigma$ (corrected by $\sigma_{in}$ and $\sigma_{th}$) of
each fit to derive the radial line-of-sight systemic velocity field and the velocity dispersion map
of 30~Dor, as shown in Fig.~\ref{velocity_map_halpha}. In the case of the radial velocity field, it was corrected by the systemic velocity of 30~Dor. Although a
single Gaussian fit does not represent the exact radial velocity of
regions that display multiple profiles, these fits still give us
important information regarding the complexity of the observed
profiles. i.e., regions that present multiple profiles will have large
values of $\sigma$ (as shown in Fig.~\ref{profile_examples}). In this
sense, both maps help us to understand the overall dynamics of 30~Dor.

In the top panel of Fig.~\ref{velocity_map_halpha} we show the
velocity field, centered on its integrated H$\alpha$ line (see
Fig.~\ref{espectro_total}). Approaching and receding regions are
represented by blue and red colors, respectively. Green regions are at
the systemic velocity of the gas in 30~Dor. In order to show the whole
kinematic behaviour of 30 Doradus, we chose a dynamic range of
70\,km\,s$^{-1}$ in the velocity map. As shown in the velocity map,
the north-eastern and south-western regions remain at roughly the
systemic velocity (258.3\,km\,s$^{-1}$, see \S~\ref{generalview}). In
the case of the south-western region, the velocity dispersion map
(lower panel of Fig.~\ref{velocity_map_halpha}) shows several narrow
H$\alpha$ profiles, especially in the neighbourhood of 30~Dor~B (in
which we find the narrowest H$\alpha$ profile detected in the FLAMES
spectra).

Another remarkable feature to note from Fig.~\ref{velocity_map_halpha}
is the presence of large expanding structures. The north-eastern edge
of shell \#2 (see Chu \& Kennicutt 1994) appears to have negative
velocities (with respect to the systemic velocity) while the cavity of
this structure appears to have positive velocities, giving a 3D view
of this shell. Several profiles within this cavity display large
values for $\sigma_{obs}$, which are the result of fitting a single
Gaussian over at least two H$\alpha$ components. In general,
Fig.~\ref{velocity_map_halpha} shows that the outer regions of 30~Dor
are at the same velocity of the integrated profile of all the spectra,
and also that the outer regions can typically be fitted by a single
Gaussian component.

\begin{figure}[t!]
\hspace{-0.1cm}
\includegraphics[width=0.83\columnwidth]{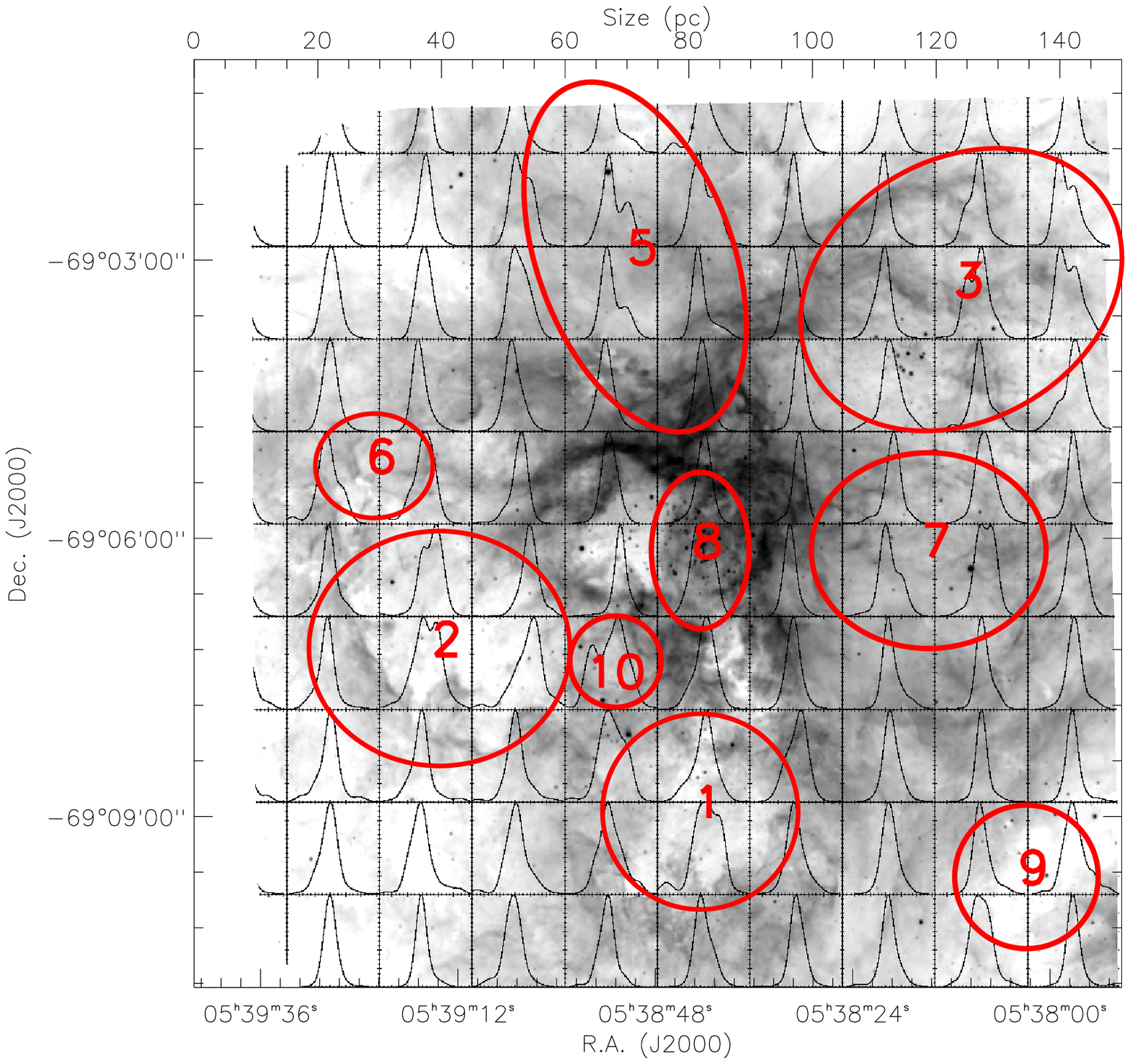}\\
\includegraphics[width=\columnwidth]{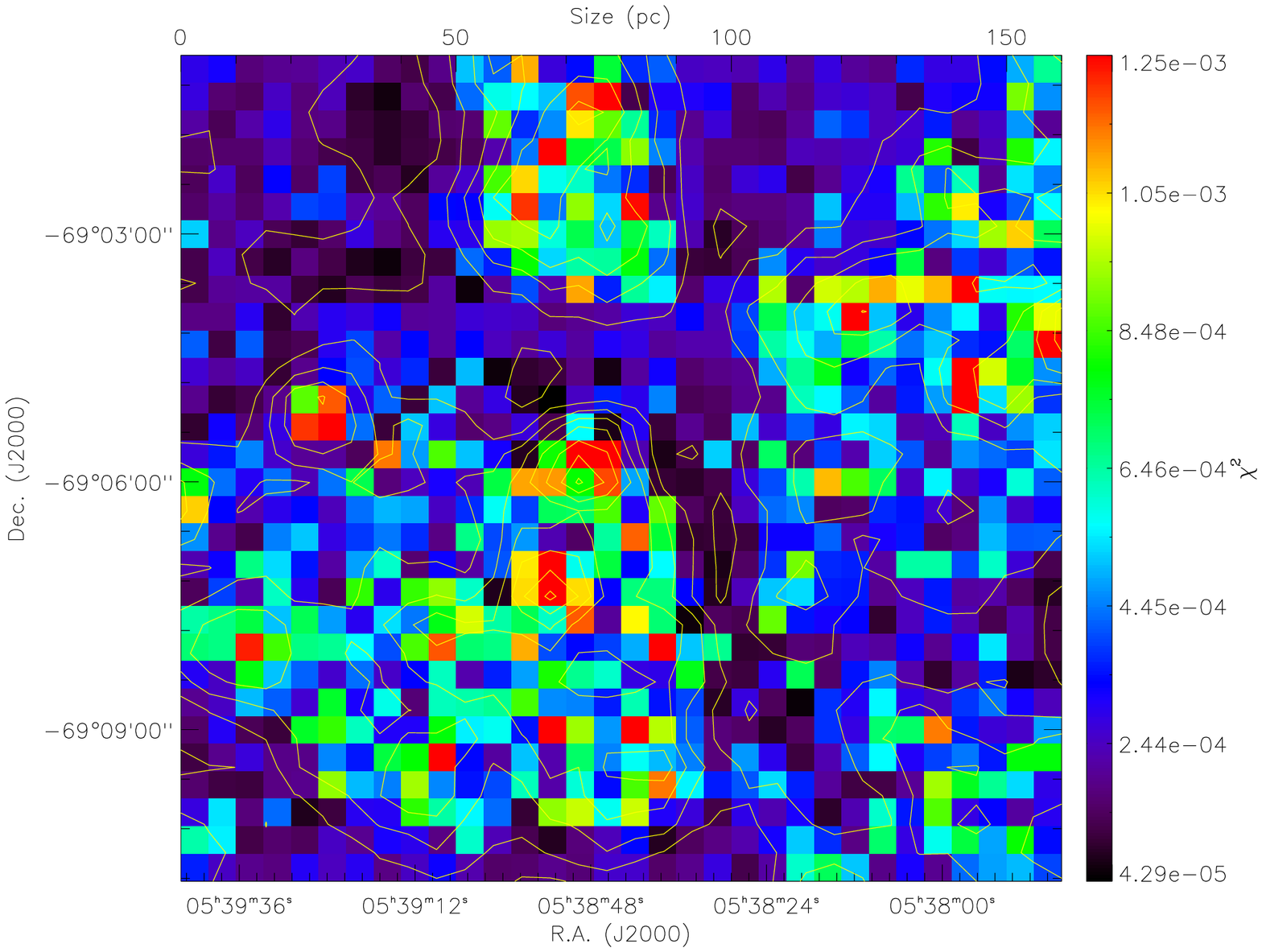}
\caption{Top panel: Large expanding structures previously catalogued in 30~Dor
  (nos. 1, 2, 3 and 5) with the five new expanding structures
  (nos. 6, 7, 8, 9 and 10) identified here. Bottom Panel: $\chi^{2}$ map for the H$\alpha$ cube of 30~Dor, derived from a single Gaussian fit on each observed profile.}
\label{shells_WCS}
\end{figure}

\subsection{Large expanding structures in 30 Doradus}
\label{expandingstructures}

We have used the H$\alpha$ datacube to search for large expanding
structures. These structures are characterized by double components in
their emission-line profiles, signature of expanding
ionized gas at these positions. To identify these structures, we have applied two methods. 
In a first instance, we have integrated the H$\alpha$ profiles shown in
Fig.~\ref{plot_adhoc_cut_Halpha_image} over regions of
1$'$\,$\times$\,1$'$, given the poor spatial sampling of our
data. Then, these integrated profiles were visually inspected in order to search for double components. In Fig.~\ref{shells_WCS}, we superimpose these integrated H$\alpha$ profiles over the H$\alpha$ image. At least ten
large expanding structures can be identified and are overlaid on the
image in Fig.~\ref{shells_WCS}.  Some of these -- shells 1, 2, 3, and
5 -- were catalogued previously (Cox \& Deharveng, 1983; Wang \&
Helfand, 1991), so we adopt the same numbers. Structures 6, 7, 8, 9 and 10 are identified for the first time by this
work.  These expanding regions present double components in their
H$\alpha$ emission, as can be noted from Fig.~\ref{shells_WCS}. 

In a second instance, the large expanding structures were searched by using the Gaussian fit on each individual profile. Given that simple H$\alpha$ profiles can be fitted with a single Gaussian, the $\chi^{2}$ value for these fits will be lower than the value presented by H$\alpha$ profiles that display double or multiple components. This fact can indicate the location of expanding structures, as profiles that are not  well fitted by a single Gaussian. In this sense, we have derived a map with the $\chi^{2}$ values obtained from a single Gaussian fit to the H$\alpha$ datacube of 30~Dor, which is shown in the bottom panel of Fig.~\ref{shells_WCS}. In this case, the Gaussian fit was performed on each observed H$\alpha$ profile, which was normalized by the total intensity of each profile. Inspecting the bottom panel of Fig.~\ref{shells_WCS}, we detect several structures that can not be fitted with a single Gaussian and that could be associated with expanding regions. Interestingly, most of these structures lie on the same position of the expanding regions detected by visual inspection. This correlation is clear for regions \#6 and \#8. Between regions \#1, \#2 and \#8, the $\chi^{2}$ map shows another peak in the $\chi^{2}$values. Inspection of the profiles at that location (top panel of Fig.~\ref{shells_WCS}) suggest the presence of a small expanding structure. Despite the analysis of the $\chi^{2}$ map is clearly more quantitative than the visual inspection of the H$\alpha$ profiles, it results necessary to combine both information in order to determine the size and the real nature of the expanding structures.

Of particular interest is structure \#6.  At this position we do not find
any star that could be producing an expansion of the ionized gas.
Studies of the diffuse X-ray emission seen by the {\it Chandra X-ray
  Observatory} appear to indicate a single-temperature thermal plasma
in this region, but with spectral signatures of both collisional
  ionization equilibrium and non-equilibrium ionization (Townsley et
al.  in preparation). This may indicate a plasma that has undergone a
recent shock and is transitioning back to ionization equilibrium,
possibly consistent with a SN remnant. This scenario is compatible
with the expanding structure found in this work; we will analyse the
main properties of these newly-detected expanding bubbles in a future
study.

\begin{figure}[t!]
\centering
\includegraphics[width=\columnwidth]{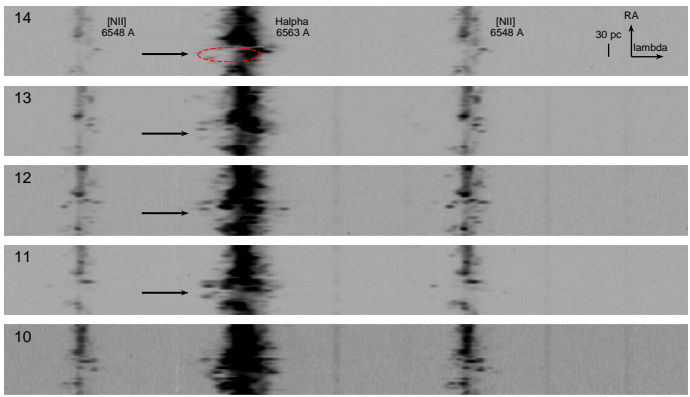}
\caption{2D cut obtained from the spectroscopic datacube of 30
  Doradus. These spectra correspond to rows 10, 11, 12, 13 and 14 from
  bottom to top (the numbers are indicated in the top left corner of each panel). The brightest lines correspond (from left-to-right)
  to [N{\sc ii}] $\lambda$6548, H$\alpha$ and [N{\sc ii}]
  $\lambda$6584. The black arrows indicate the position of an expanding
  structure. As an example, the red-dashed ellipse indicates the location of this expanding structure in row \#14.}
\label{longslit}
\end{figure}

Another method to search for expanding structures in GHRs is via
analysis of long-slit observations, e.g. the analysis in 30~Dor by Chu
\& Kennicutt (1994). To mimic their long-slit observations, we have
produced 2D cuts along selected rows of the FLAMES
datacube (i. e. at different declinations). In Fig.~\ref{longslit} we show these 2D cuts of rows 10, 11,
12, 13 and 14 of the datacube, respectively (see
Table~\ref{table_regular_all} for the RA and Dec of these positions). Row \#10 passes across the southern region of the expanding structure \#2 (see Fig.~\ref{shells_WCS}), while rows \#11, \#12, \#13 and \#14 go northwards across the structure \#2.  In these spectra, we can clearly see the H$\alpha$ and [N{\sc ii}] $\lambda\lambda$6548,\,6584 lines, with H$\alpha$ the brightest feature. By inspecting the
H$\alpha$ emission in rows 11-14 from
Fig.~\ref{longslit}, we can note the presence of an expanding
structure (as indicated by the black arrows). This structure, which can be identified
with shell \#2 in Fig.~\ref{shells_WCS}, appears as a semi-arc of emission close to the center of the H$\alpha$ line (see the red-dashed ellipse, in the case of row \#14). The expansion of this
structure can be clearly identified on the blueward side of the
H$\alpha$ line (see black arrows), but it is not easily seen on the redward side.  This
structure may be off-center with respect to the systemic H$\alpha$
emission of 30~Dor.

\section{Summary and conclusions}
\label{summary}

We have presented new VLT-FLAMES spectroscopy obtained to study the
kinematics of the ionized gas in 30~Doradus. These data consist of
regular and irregular grids of nebular observations, combined with a
stellar grid. The regular grid was combined into a datacube, allowing
us to analyse the primary kinematic features of 30~Dor, with the main
points now briefly summarised.

\begin{enumerate}

\item The kinematics of the ionized gas in 30~Dor are complex, with a
  diverse range of single and multiple H$\alpha$ profiles. In the
  brightest regions the H$\alpha$ profiles are found to be simple and
  narrow. However, the narrowest H$\alpha$ profile observed in the
  data lies close to 30~Dor~B, where a past SN explosion occurred.
  This is surprising in the sense that 30~Dor~B might be expected to
  be dominated by the kinematics of the SN remnant (i.e., multiple
  emission profiles and high-velocity components).

\item We have applied multi-Gaussian fits to the H$\alpha$ profiles of the two narrowest and broadest single-component fits detected in 30~Dor. We do not detect the
  presence of a broad, low-intensity component, as reported by Melnick
  et al. (1999) for all of their observed H$\alpha$ profiles.

\item However, the {\it integrated} H$\alpha$ profile of 30~Dor does
  display broad wings, and required Gaussian fits that include both narrow
  and broad components (as per Melnick et al.).

\item We have derived the velocity field and a velocity dispersion map
  of 30~Dor. By inspecting these maps, we found that the outer parts
  of our observed regular grid (i.e. a field-of-view of
  10$'$\,$\times$\,10$'$ centered on R136) are at the same velocity of
  the integrated H$\alpha$ profile of 30~Dor. 

\item Using the spectroscopic datacube of 30~Dor, we have identified at
  least five previously unclassified expanding structures. In one
  case (structure \#6), we did not find any star associated with
  the expanding gas at that location.

\end{enumerate}

Areas for future work include a detailed analysis of the supersonic
velocity in the integrated profile of 30~Dor (besides the origin of
the wings in this profile) and a more in-depth study of the individual
kinematic structures identified and discussed here.

\begin{acknowledgements}
  We would like to thank the referee for the useful comments that improved this paper. 
  We also acknowledge the invaluable help of Leisa Townsley in the interpretation of some
  of the expanding structures in 30~Dor. ST-F acknowledges the financial support of FONDECYT (Chile) through a
  post-doctoral position, under contract 3110087. M.R. wishes to acknowledge support from FONDECYT (Chile) grant N$^{\circ}$ 1080335. M.R. was supported by the Chilean {\sl Center for Astrophysics} FONDAP N$^{\circ}$ 15010003.

\end{acknowledgements}

\begin{appendix} 

\section{Position of the fibers}

In Table A.1 we list the fiber positioning of the regular nebular grid
(which does not include the positions of the broken fibers). In the
cases where we used a fiber from the irregular nebular grid to complete
our regular grid, we label that fiber as `IG' (irregular grid). In
Table A.2 we list the fiber positions of the irregular nebular grid.

\twocolumn
\begin{centering}
\label{table_regular_all}
\tablecaption{Fiber positions of the regular nebular grid.}
\tablehead{
\hline
ID   & X  & Y &RA (J2000)  &     DEC (J2000) & Data \\  & pix & pix &h:m:s&d:m:s & \\
\hline
}
\begin{supertabular}{cccccc}
    1  &     1   &   1 & 05:39:41.801  &  -69:10:39.900   &  Obs.  \\
    2  &     2   &   1 & 05:39:38.059  &  -69:10:39.900   &  Obs.  \\
    3  &     3   &   1 & 05:39:34.320  &  -69:10:39.900   &  Obs.  \\
    4  &     4   &   1 & 05:39:30.590  &  -69:10:39.900   &  Obs.  \\
    5  &     5   &   1 & 05:39:26.849  &  -69:10:39.900   &  Obs.  \\
    6  &     6   &   1 & 05:39:23.110  &  -69:10:39.900   &  Obs.  \\
    7  &     7   &   1 & 05:39:19.370  &  -69:10:39.900   &  Obs.  \\
    8  &     8   &   1 & 05:39:15.629  &  -69:10:39.900   &  Obs.  \\
\hline
\end{supertabular}
\end{centering}

\twocolumn
\begin{centering}
\label{table_irregular_all}
\tablecaption{Fiber positions of the irregular nebular grid.}
\tablehead{
\hline
ID   &  RA (J2000)  &  DEC (J2000) \\  &h:m:s&d:m:s\\
\hline
}
\begin{supertabular}{ccc}
 1     & 05:38:45.040  &  -69:05:09.600 \\
 2     & 05:38:47.970  &  -69:05:03.600 \\
 3     & 05:38:51.330  &  -69:04:59.900 \\
 4     & 05:38:44.980  &  -69:04:50.600 \\
 5     & 05:38:43.050  &  -69:04:31.900 \\
 6     & 05:38:55.000  &  -69:05:01.600 \\
 7     & 05:38:36.380  &  -69:06:21.200 \\
 8     & 05:38:34.580  &  -69:04:56.900 \\
\hline
\end{supertabular}
\end{centering}

\end{appendix}

\end{document}